\documentclass[fleqn]{mnras} 
\usepackage{newtxtext,newtxmath}

\usepackage[T1]{fontenc}
\usepackage{ae,aecompl}

\usepackage{graphicx}	
\usepackage{amsmath}	

\title[MW RT model]{A radiation transfer model for the Milky Way. II The global properties and large scale structure}

\author[G. Natale et al.]{Giovanni Natale$^{1}$,\thanks{E-mail: cpopescu@uclan.ac.uk}
Cristina C. Popescu$^{1,2}$, 
Mark Rushton$^{3}$,
Ruizhi Yang$^4$,
\newauthor
Jordan J. Thirlwall$^{1}$,
Dumitru Pricopi$^3$
\\
$^1$University of Central Lancashire, Jeremiah Horrocks Institute, Preston, PR1 2HE, UK \\
$^2$Max Planck Institute f\"{u}r Kernphysik, Saupfercheckweg 1, D-69117 Heidelberg, Germany\\
$^3$The Astronomical Institute of the Romanian Academy, Str. Cutitul de Argint 5, Bucharest, Romania \\ 
$^4$ University of Science and Technology of China, 230026 Hefei, Anhui, China\\}

\date{Accepted XXX. Received YYY; in original form ZZZ}

\pubyear{2016}

\begin{document}
\label{firstpage}
\pagerange{\pageref{firstpage}--\pageref{lastpage}}
\maketitle

\begin{abstract}
We obtained an axi-symmetric model for the large-scale distribution of stars and
dust in the Milky Way (MW) using a radiative transfer code that can account for the
existing near-infrared (NIR)/mid-infrared/submm all-sky emission maps of our Galaxy. 
We find that the MW has a star-formation rate of 
${\rm SFR}=1.25\pm0.2\,{\rm M}_{\odot}$/yr, a stellar mass $M_{*}=(4.9\pm 0.3)\times10^{10}\,{\rm
  M}_{\odot}$, and a specific SFR that is relatively constant with radius (except for the inner 1 kpc). We identified an inner radius $R_{\rm in}= 4.5$\,kpc beyond 
which the stellar emissivity and dust distribution fall exponentially. For
$R<R_{\rm in}$ the emissivities fall linearly towards the centre. The old stellar populations in the disk have an exponential
scalelength that increases monotonically from $h_{\rm s}^{\rm
  disk}(K)=2.2\pm 0.6$\,kpc in the NIR, to $h_{\rm
  s}^{\rm disk}(B)=3.2\pm 0.9$\,kpc at the shorter optical bands, and a scaleheight
that varies with radial distance, from $z_{\rm s}^{\rm disk}(0)
=140\pm 20$\,pc in the centre to $z_{\rm s}^{\rm disk}(R_{\odot})
=300\pm 20$\,pc
at the solar radius. The young stellar populations have a scalelength of
$h_{\rm s}^{\rm tdisk}=3.2\pm 0.9$\,kpc and a scaleheight that varies from
$z_{\rm s}^{\rm tdisk}(0)=50\pm 10$\,pc in the centre to $z_{\rm s}^{\rm
  tdisk}(R_{\odot})=90\pm 10$\,pc at the solar radius. We discovered an inner stellar
disk within the central 4.5 kpc, which we associate with the extended long bar
of the MW. Most of the obscured star formation happens within this inner
thin disk. The diffuse dust is mainly distributed in  a disk with 
scalelength $h_{\rm  d}^{\rm disk}=5.2\pm 0.8$\,kpc and scaleheight $z_{\rm
  d}^{\rm disk}=0.14\pm 0.02$\,kpc. We give the first derivation of the MW attenuation curve and present it as a functional fit to the model data. We find the MW to lie in the Green Valley of the main
sequence relation for spiral galaxies.
\end{abstract}




\section{Introduction}

The Milky Way (MW) is our nearest astrophysical laboratory for studying galaxy
formation and evolution. Yet, a good understanding of the global properties of
our Galaxy, including the total luminosity output of the different stellar
populations and their spatial distribution, the recent star-formation rate
(SFR), as well as its SF history, the total dust mass and spatial distribution 
of dust opacity, the clumpiness of the ISM, the radiation
fields, are still uncertain (see Bland-Hawthorn \&  Gerhard 2016 for a 
review on the structural, kinematic and integrated properties of the Galaxy). 
Major questions of whether the Milky Way is a typical spiral galaxy or a 
peculiar one, if its group environment and galaxy
interaction history has played a major role in shaping its global properties,
are still open and need to be addressed. This is particularly important since 
the Milky Way is de-facto the primary object used for investigations of galaxy 
evolution via studies of galactic archeology. 

Determination of the total luminosity and geometrical distribution of the
different stellar populations has been usually done combining stellar
population models with star count data: the 
Besan\c con model (Robin \& Creze 1986;  Bienayme et al. 1987; 
Robin et al. 1996, 2003), the SKY model (Wainscoat et al. 1992; 
Cohen 1993, 1994, 1995), and the TRILEGAL model (Girardi et al. 2005). 
 However, these inferences about the global distribution of stellar 
populations are limited by confusion and sensitivity of the 
surveys used to derive them, in a way sensitively depending on the galactic 
latitude due to the presence of high extinction towards the inner Galaxy. 
Great progress has been made by mitigating these limitations (Marshall et
al. 2006;  Sale 2012, 2014; Green et al. 2014, 2015; Schlafly et al. 2014) in 
surveys of higher resolutions and sensitivity, such as the Two Micron All
Sky Survey (2MASS) and PAN-STARRS (Kaiser et al. 2010) and 
are now further improved by the Large Sky Area Multi-Object Fiber Spectroscopic Telescope (LAMOST)(Cui et al. 2012)  and GAIA (Perryman et al. 2001, Gaia Collaboration 2016, 2018, Lindegren et al. 2018). For example, red clump stars surveys have been successfully  used to investigate the stellar structure of the Milky Way (e.g. Wegg et al. 2015, Clarke et al. 2019, Li et al. 2020, Sun et al. 2020, Yu et al. 2021). Nevertheless it is still difficult to derive a complete picture of the stellar emissivity of the Milky Way, in particular at large distances from the Sun, with many recent studies focusing on the anticenter and the local neighbourhood (Li et al. 2021, Gontcharov \& Mosenkov 2021). Furthermore,  
even for GAIA, and  notwithstanding the very sophisticated Bayesian techniques 
for handling selection biases (e.g. Green 2014), the effect of dust on the 
derived stellar distributions are challenging to accurately correct for, 
due to the very inhomogeneous distribution of dust (on parsec scales) in the 
diffuse interstellar medium of the galactic plane (Bovy et al. 2016).  

Therefore  
there is a need for alternative methods which are capable of deriving  
the complete distribution for all stars in the Galaxy.

It is also critical to get information about the very recent star-formation
history on time scales less than 1 Gyr, in view of theoretical predictions of
large-scale variations in the spatial pattern of this star-formation,
resulting from feedback episodes operating on timescale of several hundreds 
Million years (Tacchella et al. 2016).

Another quantity of physical importance is the distribution of dust, which is
important not only in its own right, but also because it is increasingly
recognised that dust is a good tracer of gas (Eales et al. 2012, 
Groves et al. 2015). Traditionally dust has
been inferred either from extinction measurements of stars (Lada et al. 1994,
Lombardi \& Alves 2001, Marshal et al. 2006, Lombardi 2009, Rowles \& Froebrich
2009, Schlafly et al. 2010, Berry et al. 2011, Lallement et al. 2014, 
Chen et al. 2014, Hanson \& Bailer-Jones 2014, Alves et al. 2014, Green et
al. 2014, Schlafly et al. 2014, Green et al. 2015, Lallement et al. 2018a,b, Green et al. 2019, Wang \& Chen 2019, Hottier et al. 2020, Ferreras et al. 2021) or through direct
measurements of dust emission (Reach et al. 1995, Sodroski et al. 1997,  
Schlegel et al. 1998, Finkbeiner, Davis \& Schlegel 1999, Drimmel 2000, Drimmel \& Spergel 2001, 
Planck Collaboration XI 2014, Planck Collaboration XXII 2015, Planck Collaboration X 2016, Meisner \& Finkbeiner 2015, Odegard et al. 2016). The measurements from dust extinction have 
the great advantage that one can get distances for the absorbing structures, 
since the distances of stars are known, but have the disadvantage that one 
can't probe opaque structures in this way. Measurements of dust in emission 
probe all dust in the Galaxy, but with the price that it is challenging to 
extract the geometrical distribution. 

The way to overcome the disadvantages of these previous methods while
retaining the advantages of them is to self-consistently take into account
both the extinction and emission processes, by performing a radiative transfer
(RT) calculation that follows the propagation of photons from all stellar
populations and predicts the response of dust grains to the ambient radiation
fields (Popescu 2021). To avoid the biases mentioned before about star counts the
radiative transfer methods should ideally not invoke geometrical
constraints from star
counts, but rather derive the geometrical distributions of stars and dust
through a comparison of predicted images with observed surface photometry in
both direct stellar light and dust re-radiated stellar light. 

There has been little work on radiative transfer modelling of the Milky
Way. Misiriotis et al. (2006) fitted the NIR images of the Milky Way with model
images produced with radiative transfer calculations using a description for
the distribution of stars and dust taken from Popescu et al. (2000a). However,
this study did not self-consistently link the dust emission with the
radiation fields derived from radiative transfer calculations. A fully
self-consistent model was achieved by Robitaille et al. (2012), who developed a
non axi-symmetric RT model of the Milky Way based on the SKY
model of  Wainscoat et al. (1992). The model was constrained by the 
mid-infrared (MIR) observations coming from GLIMPSE (Benjamin et
al. 2003; Churchwell et al. 2009), MIPSGAL (Carey et al. 2009), and IRAS
surveys (Miville-Desch\^{e}nes \& Lagache 2005), but not 
by observations longwards of 100 micron, which incorporate the peak of the dust
emission spectral energy distribution (SED) in the FIR and constrain the total dust luminosity,
and by observations in the submm, which constrain the dust
opacity. Furthermore, the model did not incorporate local absorption and
emission in the star forming regions, which are the main contributors
to the mid-infrared emission in star forming galaxies in the 25 and 60 micron
bands.

The Plank data allowed for the first time a good spectral and spatial coverage of the
Milky Way, presenting us with the opportunity to do a comprehensive
radiation transfer modelling of surface photometry of the Milky Way from the
NIR to the submm.
The main challenge is the lack of direct observation in the ultraviolet (UV) 
and optical range. This is a very significant problem as we know from studies 
of external galaxies that UV is not only important in heating dust around 
star-forming regions but also dust in the diffuse ISM (Popescu et al. 2002,
Hippelein et al. 2003, Popescu et al. 2005, Sauvage et al. 2005, 
Hinz et al. 2006). The second challenge is overcoming the degeneracy between 
luminosity and distance for both stellar and
dust emitting structures. One
possibility is to use radio spectroscopic observations of gas tracers and 
invoke some 
physical link between dust grains and gas in galaxies to derive the
distribution of dust. However, the transformations between the radio tracers 
and the actual distribution of the gas are themselves challenging to 
physically model and are empirically uncertain. This, in turn, may potentially 
introduce systematic error into the model predictions for the
ISRF. To overcome these challenges we use a self-consistent
radiative-transfer modelling approach in combination with
state-of-the art all-sky emission observations of the Galaxy, as
provided by the COBE, IRAS and Planck maps in the 
near-, mid-, far-infrared and submm. 

Here we present the second paper (Paper II) on a series devoted to the modelling of the Milky Way. 
In Paper I (Popescu et al. 2017) we showed the solution for the radiation fields of the Galaxy and described the implications of our model for the gamma rays produced 
via inverse Compton scattering for cosmic ray (CR) electrons, as well as for 
the attenuation of the gamma rays due to interactions with photons of the
ISRF. In this paper we present the model for the stellar and dust distribution
of the Milky Way, and describe the implications of the new model for the
broad field of galaxy formation and evolution.

The modelling of the Milky Way is part of a general effort to model the SEDs of galaxies.
Our RT model has been successful in accounting for both the spatial
and spectral energy distributions of individual galaxies 
(Popescu et al. 2000a; 
Misiriotis et al. 2001; Popescu et al. 2004; Popescu et al. 2011 - PT11;
Thirlwall et al. 2020) and in
 predicting the statistical behaviour of a variety of observables 
of the population of spiral galaxies in the local Universe 
(e.g. M\"ollenhoff et al. 2006, Driver et al. 2007, 2008, 2012; Graham \& Worley 2008; Masters et al. 2010; 
Gunawardhana et al. 2011;
Kelvin et al.  2012, 2014; Grootes et al.  2013, 2017; Pastrav et al. 2013a,b; Vulcani et 
al. 2014, Leslie et al. 2018). Grootes et al. (2014) has shown that using the RT model of PT11 to 
correct the fundamental scaling relation between specific star-formation 
rates, as measured from the UV continuum, and stellar mass for the effects of 
dust attenuation leads to a marked reduction of the scatter in this relation, confirming the ability of the PT11 model to predict the propagation 
of UV light in galaxies. Davies et al. (2016) has shown that, when 
critically compared and contrasted with various methods to derive 
star-formation rates in galaxies, the one using this RT method gives the most 
consistent slopes and normalisations of the SFR-specific stellar mass 
relations. 

The paper is organised as follows: In Sect.~\ref{sec:data} we present the COBE,
IRAS and PLANCK maps used to constrain the model and the processing of
the data. The main components of the model and its parameters are
introduced in Sect.~\ref{sec:model}, while the radiative transfer codes are described
in Sect.~\ref{sec:codes}. The optimisation procedure and the main steps taken in
fitting the NIR/FIR/submm images of the Milky Way are described in
Sect.~\ref{sec:fitting}. The results for the global properties of the Milky Way are
given in Sect.~\ref{sec:results}. In the same section we also give the results for the
spatial distributions of stars and dust. In Sect.~\ref{sec:discussion} we 
discuss our predictions regarding  the spatial variation of different physical quantities (e.g. SFR, stellar mass). We also make predictions for the attenuation curve of the Milky Way.
In Sect.~\ref{sec:comparison} we compare the
solution obtained for the Milky Way with solutions derived for
external galaxies. We give the summary and conclusions in Sect.~\ref{sec:summary}.

\section{Data and emission strips}
\label{sec:data}

\begin{figure*}
\centering
\includegraphics[scale=0.5]{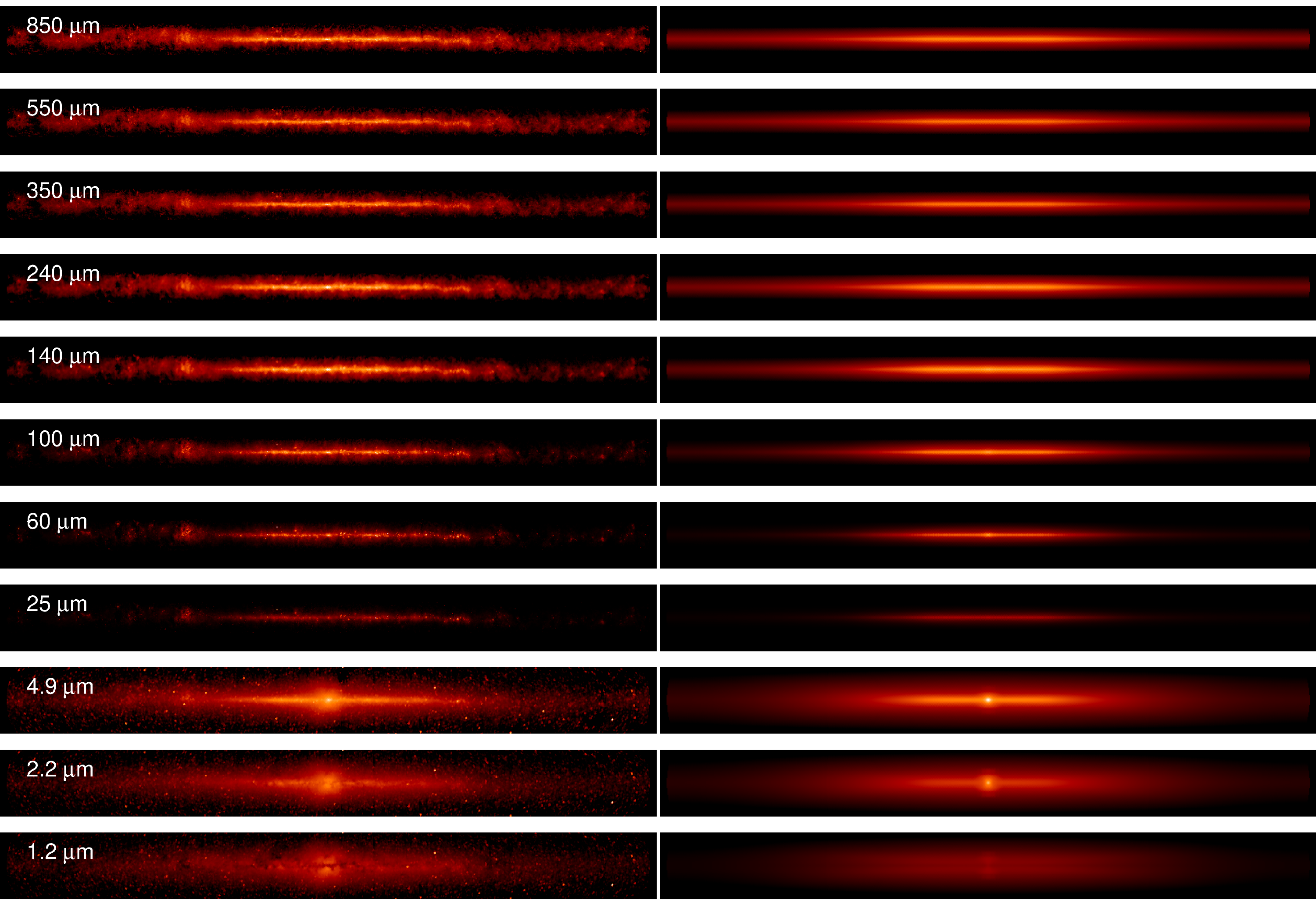}
\caption{Observed (left) and model (right) background-subtracted
  Galactic Plane Strip (as defined in Sect.~\ref{sec:data}) maps of
  the Milky Way. We note that we did not try to reproduce the complex peanut/boxy shape of the bulge, but instead we used a simple de Vaucouleurs bulge in the model.} 
\label{fig:map_grid}
\end{figure*}
\begin{figure*}
\centering
\includegraphics[scale=0.8]{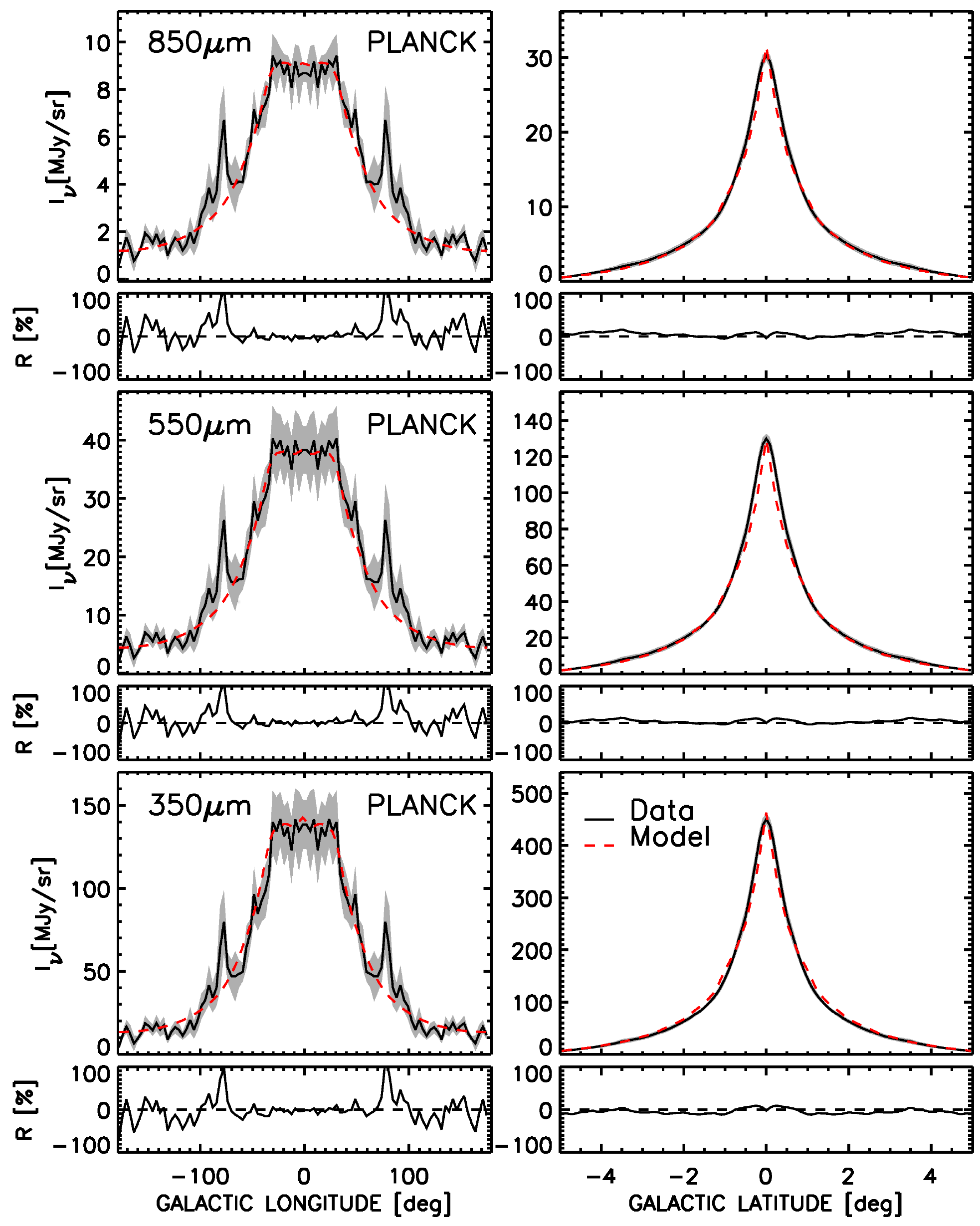}
\caption{Left: Longitude profiles averaged over latitude and mirrored (between the clock-wise and anti clock-wise directions with respect to the Galactic Centre direction). Because of the mirroring we present the profiles with the x-axis as $(-180^{\circ},0^{\circ})$ instead of $(180^{\circ},360^{\circ})$. Right: Latitude profiles averaged over longitude and mirrored with respect to the Galactic Plane. The profiles are derived from the Galactic Plane strip (see Sect.~\ref{sec:data}) and are plotted at wavelengths in the submm-FIR range, with continuum black-line for the observations and dashed red line for the model.
 The shaded area represents the uncertainties in the observed
  profiles, derived as described in Sect.~\ref{sec:data}.} 
\label{fig:dust_profiles_1}
\end{figure*}
\begin{figure*}
\centering
\includegraphics[scale=0.8]{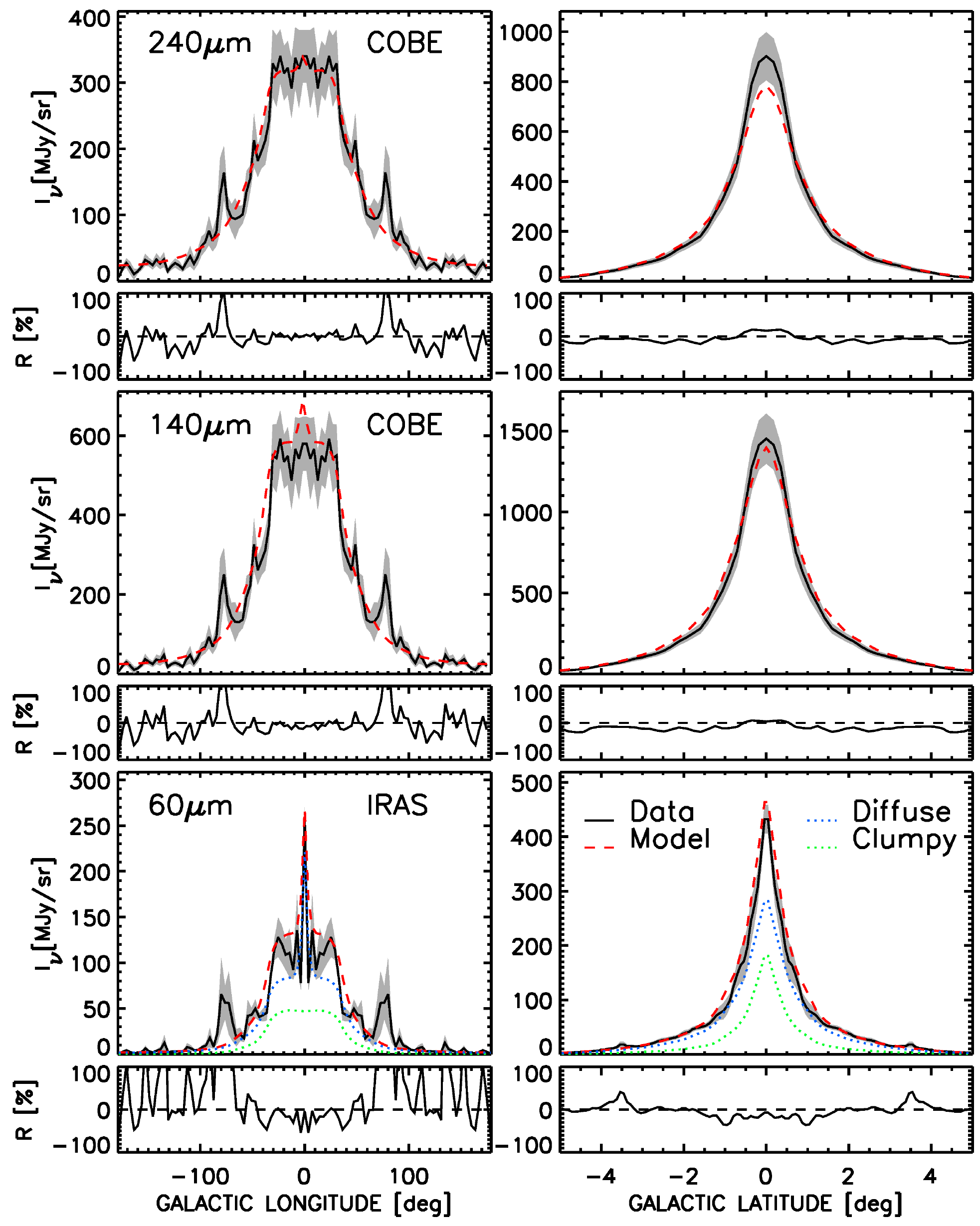}
\caption{As in Fig.~\ref{fig:dust_profiles_1}, but for wavelengths in the FIR-MIR range. In the MIR we also plot the contribution of the diffuse (blue dotted-lines) and clumpy (green dotted-lines) components to the dust emission profiles.} 
\label{fig:dust_profiles_2}
\end{figure*}
\begin{figure*}
\centering
\includegraphics[scale=0.8]{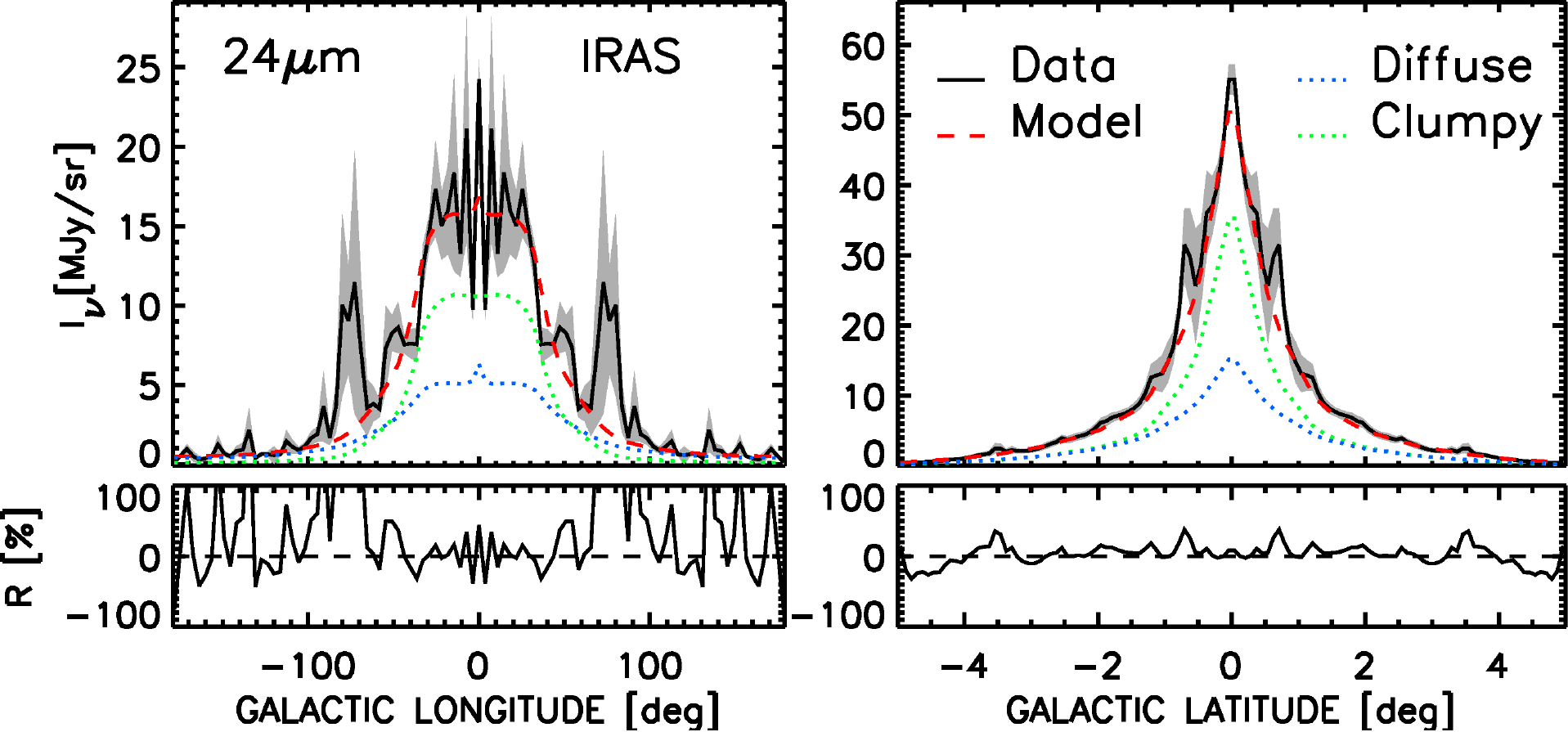}
\caption{As in Fig.~\ref{fig:dust_profiles_1}, but at $24\mu$m. We also plot the contribution of the diffuse (blue dotted-lines) and clumpy (green dotted-lines) components to the dust emission profiles.} 
\label{fig:dust_profiles_3}
\end{figure*}
\begin{figure*}
\centering
\includegraphics[scale=0.8]{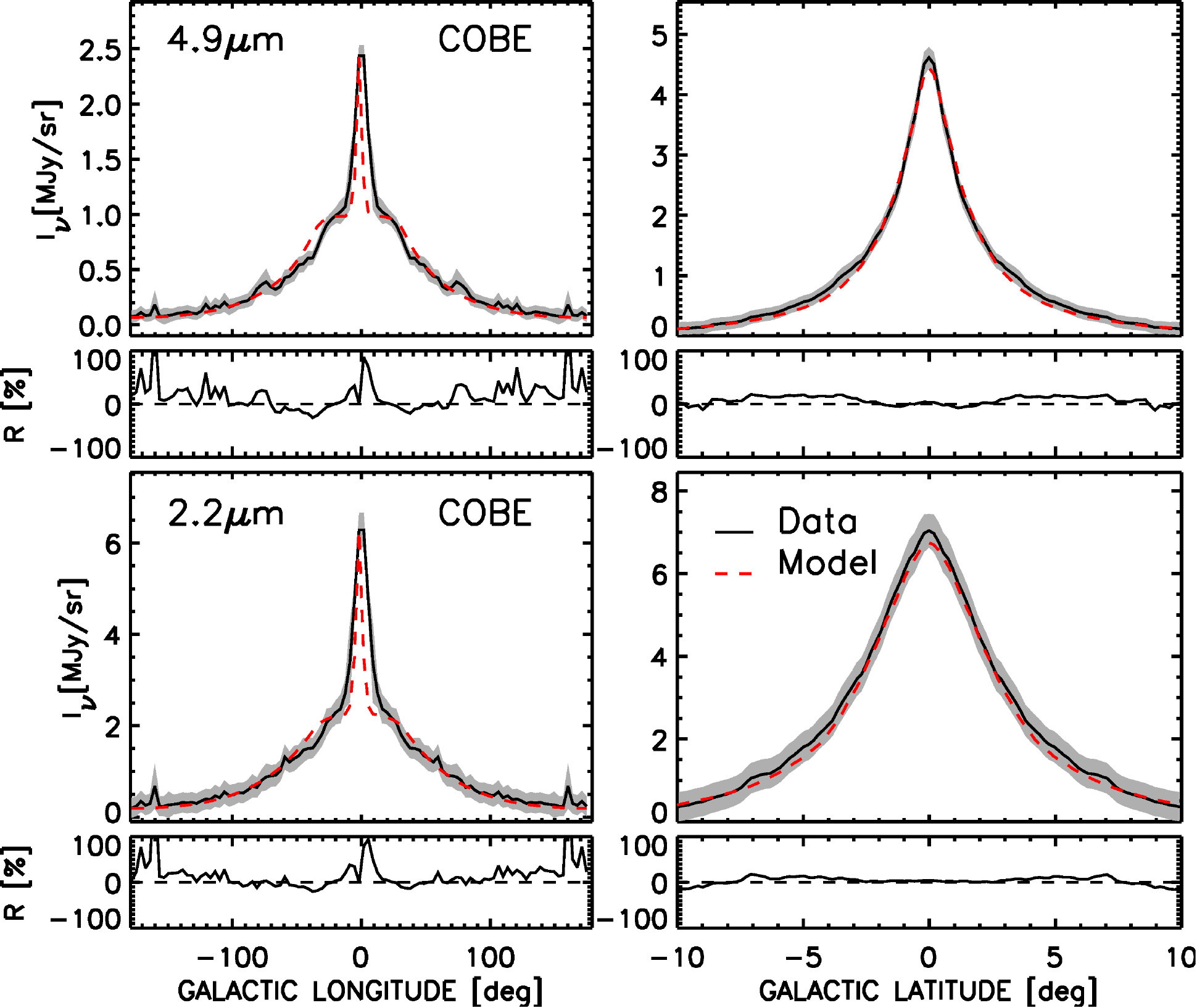}
\caption{As in Fig.~\ref{fig:dust_profiles_1}, but in the NIR.} 
\label{fig:dust_profiles_nir}
\end{figure*}
\begin{figure*}
\centering
\includegraphics[scale=0.8]{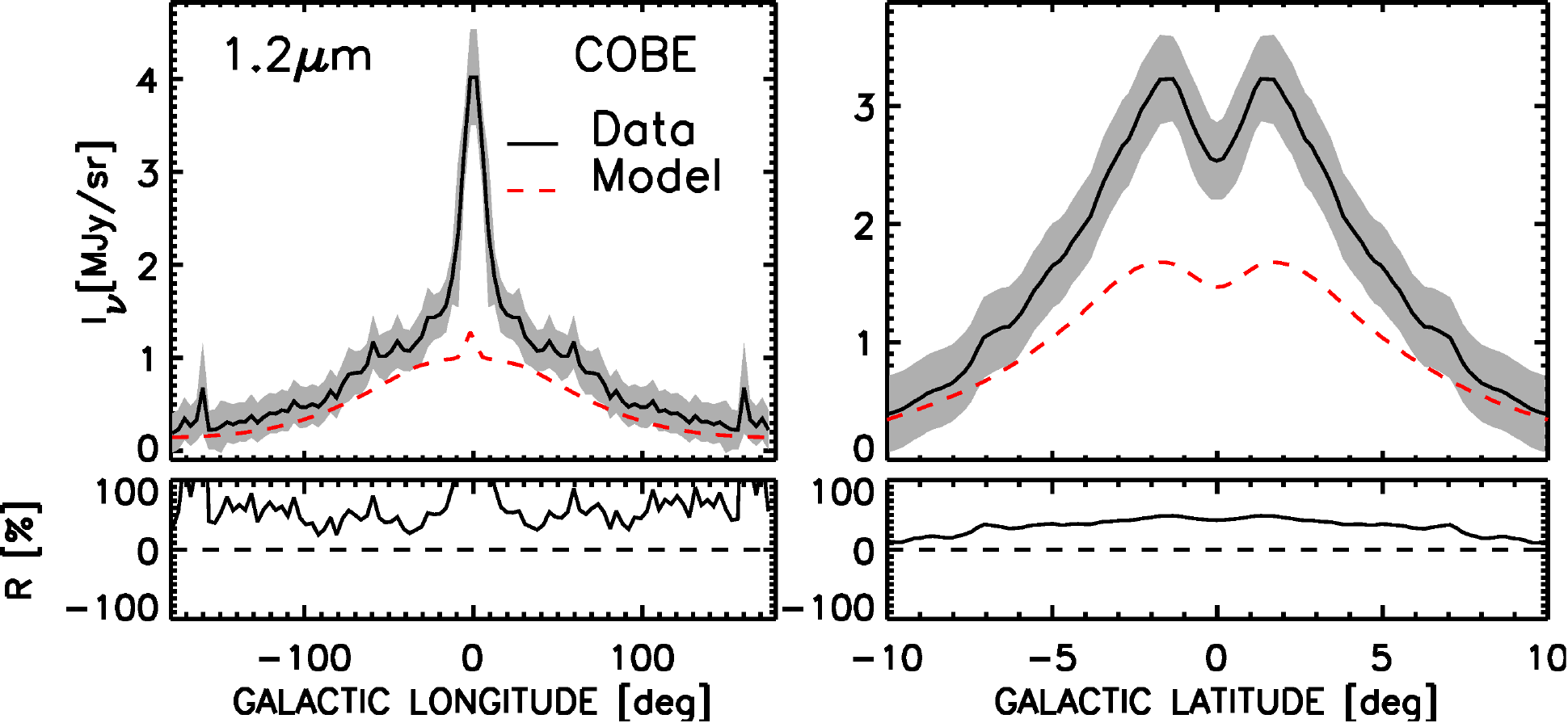}
\caption{As in Fig.~\ref{fig:dust_profiles_1}, but in the J band.  We note that we did not try to reproduce the complex peanut/boxy shape of the bulge, but instead we used a simple de Vaucouleurs distribution. In particular at this band where the attenuating effects of the dust start to play a role, the interplay between dust attenuation and a more complex stellar distribution means that our simple model of the bulge cannot reproduce the data.} 
\label{fig:j_profile}
\end{figure*}

The data we used in this work include full-sky zodiacal-light subtracted maps from COBE, IRAS and PLANCK. Specifically, we included the bands centred at 1.2, 2.2, 3.5, 4.9, 140 and 240\,$\mu$m from COBE DIRBE (downloaded from the CADE database; Paradis et al. 2012)\footnote{http://cade.irap.omp.eu/dokuwiki/doku.php?id=cobe}; at 12, 25, 60 and 100\,$\mu$m from IRAS (IRIS reprocessing\footnote{http://www.cita.utoronto.ca/~mamd/IRIS/IrisOverview.html}, Miville-Desch\^{e}nes
\& Lagache 2005); and at 350, 550 and 850\,${\mu}$m from PLANCK High
Frequency Instrument (see Ade et al. 2014 and Adam et al. 2015, maps downloaded from NASA/IPAC Infrared Science Archive\footnote{http://irsa.ipac.caltech.edu/data/Planck/release\_1/all-sky-maps/}). From the 850\,$\mu$m map, we subtracted the contribution from the CO J = 3 -> 2 line using the CO emission maps provided by the Planck collaboration (Planck Collaboration 2014,
paper XIII). We also masked the central 1\, deg square around the Galactic Centre in the maps from 140 to 850$\mu$m, because of the presence of a bright source not included in our model.

The observed all-sky FIR maps are known to display a pattern of irregular dust
emission structures, particularly visible at Galactic latitudes higher than
a few degrees, and most probably made of predominantly nearby Cirrus clouds.
In order to avoid contamination from these nearby structures, as well
as from extragalactic sources, we limited the comparison between data
and model maps within a strip of fixed size in latitude, centred
around the Galactic Plane. This procedure also eliminates any
possible Galactic dust emission halo contribution from the data,
should such an emission component exist and be important.  
We refer to the strip centred around the Galactic Plane as the \lq\lq
Galactic Plane Strip\rq\rq. Its size is
$\pm$5\, deg for the wavelengths from 12 to 850\,$\mu$m
(dominated by dust emission). 
In the NIR bands, between 1.2 and 4.5\,$\mu$m, where the emission is
dominated by stellar emission, a larger strip in latitude
was needed, of $\pm$15\, deg, in order to include the bulge emission and the emission from the old stellar populations belonging to the disk, which becomes more vertically extended at large radii. 
In order to subtract the background emission from the Galactic Plane Strip and take
into account the background variability with longitude, we estimated the
background in regions of 2\,deg in latitude above and below the
Galactic Plane Strip
and at regular 1\,deg intervals in longitude. Then, a linear function
was used to fit the background for each bin in longitude. In this way, we
subtracted the background separately for each set of pixels located
within each of these narrow longitude intervals.   Examples of
background-subtracted Galactic Plane Strip  maps derived from observations at various wavelengths are given in the left
panels of Fig.~\ref{fig:map_grid}.    

The Galactic Plane Strip maps were used to produce averaged longitude and latitude profiles. The longitude profiles were 
averaged over latitude and mirrored (between clock-wise and anti clock-wise directions with respect to the Galactic
Centre direction). Because of the mirroring we redefine the longitudes within the $(180^{\circ},360^{\circ})$ range as negative longitudes $(-180^{\circ},0^{\circ})$. The latitude profiles were averaged over longitude and mirrored with respect
to the Galactic Plane (averaged between positive and negative
latitude). These observed averaged profiles  were used to fit the
axi-symmetric model of the Galaxy. Examples
can be seen in Figs.~\ref{fig:dust_profiles_1} - \ref{fig:j_profile}.       

The errors in the derived averaged surface brightnesses have been
calculated by taking into account the calibration errors, the
background fluctuations and the configuration noise (arising from
deviations of the observed brightness from an axisymmetric
distribution). These derivations are detailed in Appendix~\ref{sec:proferrors} and the corresponding errors are plotted as grey shades surrounding the observed profiles from Figs.~\ref{fig:dust_profiles_1} - \ref{fig:j_profile}. Overall the errors are dominated by the configuration noise, although the weight of the different sources of error vary largely with wavelength, and position in the longitude/latitude profile.

\section{Model description}
\label{sec:model}

Our model is based on the axisymmetric RT model of PT11 for the UV to submm 
emission of external galaxies, in which the geometry of dust opacity and 
stellar emissivity is prescribed by parameterised analytic functions. 

While retaining this overall formalism, an optimisation was
performed for the geometrical parameters of
the morphological components of the MW, based on the detailed surface brightness
photometry available for our Galaxy, in particular on Planck data. At
the same time we also had to implement a new methodology that
deals with the inner view of a galaxy and with the lack of direct observational
constraints in the UV-optical regime within the solar circle.

Thus our model of the Milky Way contains the stellar and dust components
from the generic model of PT11, the old stellar disk (the disk) and associated 
dust (the dust disk), the young stellar disk (the thin disk) and associated 
dust (the thin dust disk), the stellar bulge and the clumpy
component. 
 
The
terminology used here was motivated by the vertical extent of the different stellar and
dust/gas components, starting with the thinner structure, that of the 
molecular layer of a  galaxy where young stars form, which is known to 
have a scaleheight in the range 50-90 pc.
This is usually called the thin disk by the community working on molecular gas
measurements and star-formation and this was also the terminology
adopted when we first introduced the modelling technique in Popescu et
al. (2000a), and which we continue to use in this paper. Thus, we
describe our model as having a thin disk 
(scalelength up to 90 pc) and a disk (scalelength up to a few hundered pc). 
There is however another astrophysics community, looking at galaxy haloes
and extraplanar disks, or working on N-body/SPH simulations for
galaxies, which refers to a ``ultra thin disk'' and a ``thin disk'', respectively
(also used in the review of the Milky Way by Bland-Hawthorne \& Gerhard 2016). We continue to use our previous terminology, but
draw the attention to the reader of the variations in terminology found in the field.

In addition to the stellar and dust components used in the generic model of
PT11, we found that for the Milky Way it was neccesary to introduce an inner
stellar component, refered to as the ``inner thin disk'', and
to alter the exponential behaviour of the surface brightness distribution in 
the centre of the disk components. Thus in our model of the Milky Way the 
stellar volume emissivity and the dust density distributions for all the disk 
components $i$ are described by the following general formula: 

\begin{equation}
\label{eq:model}
j_{\rm i} (R,z) = \begin{cases} 
{\displaystyle
A_o\left[\frac{R}{R_{\rm in}}(1-\chi) + \chi\right]
\exp\left(-\frac{R_{\rm in}}{h_{\rm i}}\right) T_{\rm i}}\\ 
 
\hspace{4.7cm} {\rm if}\hspace{0.1cm} R < R_{\rm in}\\\\
{\displaystyle
A_o
\exp{\left(-\frac{R}{h_{\rm i}}\right)T_{\rm i}
}
\hspace{1.6cm}{\rm if}\hspace{0.1cm}
R_{\rm in} \leq R \leq R_{\rm t}
}
\end{cases}
\end{equation}

\noindent
with: 
\begin{equation}
\label{eq:TRz}
T_{\rm i} = \frac{z_{\rm i}(0)}{z_{\rm i}(R)}\rm{sech}^2{\left(\frac{z}{z_{\rm i}(R)}\right)}
\end{equation}
\begin{equation}
\label{eq:chi}
\chi = \frac{j_{\rm i}(0,z)}{j_{\rm i}(R_{\rm in},z)}
\end{equation}
and
\begin{gather}
\label{eq:flare1}
z_i(R) = z_i(0) + \left(z_i(R_{\rm in})-z_i(0)\right)\left(\frac{R}{R_{\rm
    in}}\right)^\gamma \\ 
\label{eq:flare2}
\gamma = \log\left({\frac{z_i(R_\odot)-z_i(0)}{z_i(R_{\rm in}) - z_i(0)}}\right)/\log\left({\frac{R_\odot}{R_{\rm in}}}\right)
\end{gather}

\noindent
where
$R$ and $z$ are the radial and vertical coordinates, $h_i$ is the scale length, $z_i(R)$ is the scale height dependent on the radial
distance $R$, $A_o$ is a constant determining the scaling of
$j_{\rm i}(R,z)$, $\chi$ is a parameter describing the linear slope of the
radial distributions interior to an inner radius $R_{\rm in}$, 
$R_\odot$ is the radial distance of the Sun to the galactic center,
assumed here (and in Paper I) to be $R_\odot = 8$\,kpc, and $R_{\rm t}$ is the truncation
radius of the exponential distribution. As in Paper I we also assume $z_\odot=0$. In principle $\chi$ and $R_{\rm in}$ should also carry an index \lq\lq i\rq\rq\,, but because these parameters were found to be the same for all dust and stellar components of the MW model, we omit their index \lq\lq i\rq\rq.

We anticipate that the particular
shape for $j_{\rm i}(R)$ at short radii, deviating from the exponential function, has been motivated by the impossibility of 
reproducing the flat shape of the observed average surface brightness as a 
function of longitude using the former function. Instead, as it will be seen 
in Sect.~\ref{sec:fitting}, the linear decrease at low radii
allows to reproduce the observed profiles. 
 
Similarly, the less cuspy latitude profiles of the observed images were better reproduced
by a $j_{\rm i}(z)$ following a sech2 law rather than an exponential, and
the variation of the latitude profiles along the latitude required the
introduction of  a flare of the vertical distribution, by considering a
general expression for $z_i(R)$ as given in
Eqs.~\ref{eq:flare1} and ~\ref{eq:flare2}. 

It is well known that the Milky Way has a complex boxy/peanut bulge (see Bland-Hawthorne \& Gerhard 2016, Wegg et al. 2015). However, for the purpose of our axi-symmetric model we implement in this paper an ellipsoidal bulge described by a Sersic distribution. The implications of this simplication will be discussed later in the paper. Thus,   we used a Sersic distribution whose stellar volume 
emissivity $j_\nu(R,z)$  is defined as:   

\begin{equation}
\label{eq:bulge}
j(R,z) = j(0,0)\, 
\sqrt{\frac{b_{\rm s}}{2\pi}}\,\frac{(a/b)}{R_{\rm e}}\,
\eta^{(1/2n_{\rm s})-1}
\exp{(-b_{\rm s}\, \eta^{1/n_{\rm  s}})}
\end{equation}

\noindent
with:
\begin{equation}
\label{eq:bulge1}
\eta(R,z) = \frac{\sqrt{R^2 + z^2(a/b)^2}}{R_{\rm e}}
\end{equation}
where $b/a$ is the axis ratio, $R_{\rm e}$ is the
effective radius and  
$b_{\rm s}$ is a constant depending on the value of the Sersic
index $n_{\rm s}$: 
\newpage
\begin{equation}
\label{eq:bulge2}
b_{\rm s} = 
\begin{cases}
{1.67835\hspace{0.2cm} {\rm for}\hspace{0.2cm} n_{\rm s}=1}\\
{3.67206\hspace{0.2cm} {\rm for}\hspace{0.2cm} n_{\rm s}=2}\\
{5.67017\hspace{0.2cm} {\rm for}\hspace{0.2cm} n_{\rm s}=3}\\
{7.66925\hspace{0.2cm} {\rm for}\hspace{0.2cm} n_{\rm s}=4}\\
{9.66872\hspace{0.2cm} {\rm for}\hspace{0.2cm} n_{\rm s}=5}\\
{11.6684\hspace{0.2cm} {\rm for}\hspace{0.2cm} n_{\rm s}=6}\\
{13.6681\hspace{0.2cm} {\rm for}\hspace{0.2cm} n_{\rm s}=7}\\
{15.6679\hspace{0.2cm} {\rm for}\hspace{0.2cm} n_{\rm s}=8}\\
{17.6678\hspace{0.2cm} {\rm for}\hspace{0.2cm} n_{\rm s}=9}\\
{19.6677\hspace{0.2cm} {\rm for}\hspace{0.2cm} n_{\rm s}=10}
\end{cases}
\end{equation}

The integration of the model distributions (Eqs.~\ref{eq:integral},\ref{eq:TR},\ref{eq:Tz})  provides the total
luminosity spectral density $L^{i}(\lambda)$ (Eq.~\ref{eq:lum}), if these refer to the
stellar components, or the total dust mass $M_{\rm dust}$
(Eq.~\ref{eq:mass}), if the
  distributions describe the dust components. For a fixed geometry,
  the stellar luminosity  is
  proportional with the amplitude of the model distribution (central
  volume luminosity density) and because of
  this we refer to this as to amplitude parameters. For the dust
  distribution we prefer to use the central face-on dust opacity
  (Eq.~\ref{eq:density}) as the amplitude parameter.

In the following we clarify the functional shapes and properties of
each component. 

\subsection{Stellar components} 

\subsubsection{The disk} 

The disk component containing the old stellar populations and emitting 
preferentially in the optical and NIR is described by the geometrical 
parameters $h_s^{\rm disk}$, $z_s^{\rm disk}$, 
$R_{\rm in, s}^{\rm disk}$ and $\chi_s^{\rm disk}$ and the amplitude parameters
$L^{\rm disk}$ ($\lambda$). At the
wavelengths available to observations (J, K, L, M bands), the values of these 
parameters are constrained from data, as described in
Sect.~\ref{sec:fitting}. For the optical regime, where no information is
available, we assumed that the scalelength  $h_s^{\rm disk}$ 
increases with decreasing wavelength in the same ratio to the K band 
scalelength (which is contrained from data) as in the generic model of PT11 
(their Table E.1). By the
same token, the scale-height $z_s^{\rm disk}$ was fixed from PT11 to 
be the same at all wavelengths, assumption that was successfully tested to be 
correct for the available observations in J, K, L, M 
(see Sect.~\ref{sec:fitting}). In 
addition, the parameters 
$R_{\rm in, s}^{\rm disk}$ and $\chi_s^{\rm disk}$ were also found to be independent
of wavelength, and therefore fixed to the values derived from the available 
observations. The SED of the intrinsic stellar emissivity 
in the B,V,I was assumed to have the shape (color) of the fixed template from 
Table~E.2 in PT11, and was scaled to the amplitude of the SED constrained from 
observations in the NIR.

\subsubsection{The thin disk}
\label{sec:thin_disk}

The thin disk component containing the young stellar populations dominates the 
output in the UV and is described by the geometrical parameters 
$h_s^{\rm tdisk}$, $z_s^{\rm tdisk}$, $R_{\rm in, s}^{\rm tdisk}$ and 
$\chi_s^{\rm  tdisk}$  and the amplitude parameters
$L^{\rm tdisk}$. Since for the young stellar populations there are no direct
observational constraints, the value of these parameters were constrained from
the dust emission data, as described in Sect.~\ref{sec:fitting}, under the
assumption that $h_s^{\rm tdisk}$ and $z_s^{\rm tdisk}$ do not vary with
wavelength, as in PT11, and that the color of the SED of the intrinsic stellar
emissivity is that given in Table~E.2 from PT11. For the fixed colour of the
SED, the total luminosity of the young stellar disk $L^{\rm tdisk}$ is
expressed in terms of a star-formation rate ${\rm SFR}^{\rm tdisk}$, using Eqs.~16, 17 and
18 from PT11. As in our previous modelling, we prefer to use ${\rm SFR}^{\rm tdisk}$ as
the amplitude parameter instead of $L^{\rm tdisk}$.

\subsubsection{The inner thin disk}

The inner thin disk component is described by the geometrical parameters 
$h_s^{\rm in-tdisk}$, $z_s^{\rm in-tdisk}$, $R_{\rm in, s}^{\rm in-tdisk}$ and 
$\chi_s^{\rm  in-tdisk}$ and the amplitude parameters $L^{\rm
  in-tdisk} (\lambda)$. The
inclusion of this additional stellar component was motivated by the observed
data in both stellar and dust emission. The value of the corresponding
parameters were therefore constrained from data as described in 
Sect.~\ref{sec:fitting}.

\subsubsection{Bulge}

The bulge of the Milky Way is known to have a rather peculiar shape (boxy/peanut shape)(Bland-Hawthorne \& Gerhard 2016, Wegg et al. 2015) and, in addition, a bar component tightly connected to the bulge structure (Martinez-Valpuesta \& Gerhard 2011, Romero-Gomez et al. 2011, Wegg \& Gerhard 2013, Wegg et al. 2015). This clearly non axis-symmetric feature cannot be reproduced in detail by our simple description for the bulge volume emissivity. In this work we used the  Sersic distribution with $n_s=4$ (Eqns.~\ref{eq:bulge}, \ref{eq:bulge1}, \ref{eq:bulge2}), which, although imperfect, gives some overall description of the average longitude and latitude surface brightness profiles at most bands. 

\subsection{Dust components}

\subsubsection{The dust disk}

The dust disk is one of the main components of our generic model from
PT11, and it describes the large scale distribution of diffuse dust
associated with the bulk of the stellar population in a galaxy and
with the HI gas. Its main characteristic is a scaleheight $z_d^{\rm
  disk}$  that is smaller than that of the old stellar populations
$z_s^{\rm disk}$, but still larger than that of the young stellar
populations $z_s^{\rm tdisk}$. Another feature is  a scalelength $h_d^{\rm  disk}$  that is
larger than that of the old stellar disk $h_s^{\rm  disk}$ . These characteristics have
been first derived from modelling edge-on galaxies by Xilouris et
al. (1997, 1998, 1999), and have been used and shown to account for the panchromatic
modelling of edge-on galaxies in Popescu et al. (2000a), Misiriotis et al. (2001) and Popescu
et al. (2004), and adopted in our generic model of PT11. Further studies
made by other groups have also confirmed these characteristics
(Bianchi \& Xilouris 2011, Schechtman-Rook et al. 2012, De Geyter et
al. 2013, 2014). As we will show in this paper, these characteristics
are found to be exhibited by the dust disk of the Milky Way as
well. As with the stellar disks, the geometrical parameters of the
dust disk are  $h_d^{\rm disk}$, $z_d^{\rm disk}$, 
$R_{\rm in, d}^{\rm disk}$ and $\chi_d^{\rm disk}$. The amplitude
parameter is the B-band central face-on opacity $\tau^{\rm f, disk}(B)$.

\subsubsection{The thin dust disk}

The thin dust disk is a generic feature of the PT11 model, and
represents the diffuse dust associated with the young stellar
population. This dust was fixed in PT11 to have the same scalelength $h_d^{\rm tdisk}$ and
scaleheight $z_d^{\rm tdisk}$ as for the young stellar disk,
assumption that is kept in the modelling of the Milky Way. The
geometrical parameters of the thin dust disk are  $h_d^{\rm tdisk}$, $z_d^{\rm tdisk}$, 
$R_{\rm in, d}^{\rm tdisk}$ and $\chi_d^{\rm tdisk}$. The amplitude
parameter is the B-band central face-on opacity $\tau^{\rm f, tdisk}(B)$.

\subsubsection{Clumpy component}

Another generic feature of the PT11 model which we preserve here is
the clumpy component, representing the emitting dust in
the vicinity of young star-formation regions. The clumps have small
filling factor in our model, such that they do not affect
significantly the light propagating on kpc scales, 
but they block efficiently the light from young stars inside the clouds. The absorbed luminosity is then re-emitted strongly in
the mid-infrared where this component dominates the observed total
emission. In our generic model the clumpy component was assumed to
follow the same distribution as that of the young stellar disk. For
the Milky Way, however, we found that the clumpy component is not so
extended as the young stellar disk, but rather follows the same
distribution as the inner thin disk. 
The clumpy component is described by the amplitude parameter $F$, 
which was defined in Popescu et al. (2000a) and Tuffs et al. (2004) to represent the fraction of the total luminosity
of massive stars locally absorbed in star-forming clouds (see
Sect. 2.5.1 from PT11 for a detailed description of the escape
fraction of stellar light from the clumpy component).

All the geometrical parameters of our model are listed in Table~\ref{tab:geom} (the free parameters) and Table~\ref{tab:geom_fix} (the fixed parameters).
\begin{table}
\caption{The geometrical parameters of the model that are constrained from 
data. All the length parameters are in units of kpc.}
\label{tab:geom}
\begin{tabular}{ll}
\hline\hline
$R_{in}$                          & 4.50$\pm$0.03\\
$\chi$                            & 0.5$\pm$0.1\\
$h_{\rm s}^{\rm disk}(J,K,L,M)$           & (2.20, 2.20, 2.6, 2.6) $\pm$0.6\\
$z_{\rm s}^{\rm disk}(0, R_{in}, R_{\odot})$ & (0.14, 0.17, 0.30)$\pm$0.02\\
$h_{\rm s}^{\rm tdisk}$            & 3.20$\pm$0.9\\
$h_{\rm s}^{\rm in-tdisk}$            & 1.00$\pm$0.3\\
$z_{\rm s}^{\rm in-tdisk}(0, R_{in}, R_{\odot})$  & (0.05, 0.067, 0.09) $\pm$0.01\\
$h_{\rm d}^{\rm disk}$             & 5.2 $\pm$0.8\\
$z_{\rm d}^{\rm disk}$             & 0.14$\pm$0.02\\
$R_{eff}$                               & 0.4$\pm$0.08\\
$b/a$                                  & 0.6\\
\hline
\end{tabular}
\end{table}

\begin{table}
\caption{The geometrical parameters of the model that are fixed from 
PT11 or from other considerations. All the length parameters are in units of kpc.}
\label{tab:geom_fix}
\begin{tabular}{ll}
\hline\hline
$h_{\rm s}^{\rm disk}(B,V,I)$       & 3.20, 3.10, 2.80\\
$z_{\rm s}^{\rm tdisk}(0, R_{in}, R_{\odot})$  & 0.05, 0.067, 0.09\\
$h_{\rm d}^{\rm tdisk}$            & 3.20\\
$z_{\rm d}^{\rm tdisk}(0, R_{in}, R_{\odot})$ & 0.05, 0.067, 0.09\\
$R_{\rm t}$                         & 14.\\
$n_s$ & 4\\
\hline
\end{tabular}
\end{table}

\section{The radiative transfer codes}
\label{sec:codes}

For the purpose of finding a solution for the stellar emissivity and
dust distribution of the Milky Way we used both the radiative transfer
code from Popescu et al. (2011), which utilises a modified version of the Kylafis \& Bahcall (1987)
code, and the DARTRay code (Natale et al. (2014, 2015,
2017). The Kylafis \& Bahcall (1987) code employs a ray-tracing algorithm
and the method of scattered intensities, introduced by Kylafis \&
Bahcall (1987) in an implementation by Popescu et al. (2000a), which
(as in the original implementation of Kylafis) avoids obvious pitfalls
highlighted by Lee et al. (2016), while preserving speed and accuracy,
as demonstrated in Popescu \& Tuffs (2013) and Natale et
al. (2014). The DARTRay is a ray-tracing code that provides an explicit calculation of all
orders of scattered light. However, DART-Ray does not use a
brute-force ray-tracing algorithm, but takes advantage of the fact that the
radiation sources within a model do not contribute significantly to
the radiation field energy density everywhere but only within a fraction of it called the source
influence volume. DARTRay estimates the extent of the source influence
volumes and performs radiation transfer calculations only within
them. As shown in Natale et al. (2017), in dusty objects the extent
of this volume could be quite reduced relative to the size of a model,
especially for the scattered light sources, which are low intensity
sources compared to the sources actually producing radiation, such as
stars and dust thermal emission. The efficiency of the latest version
of the DARTRay code has been tested in Natale et al. (2017) for the
Milky Way model presented here. Because the model developed in this
work is an axi-symmetric model, we use the 2D mode of the DART-Ray
code, which is about a factor of 8 times faster than the standard 3D
mode.

While most of the optimisation has been done using the Popescu et
al. (2011) code, results have been checked running both codes. In
addition, the surface brightness maps, as seen by an observer within the RT model, have been produced with DART-Ray.
The output is in HEALPix format, which is a format used in all-sky
surveys, including the Planck data used in this work. 

In the mid-infrared the model maps had to be calculated using dust
self-absorption, and for this reason the DART-Ray code was used to
derive them. In the FIR/submm the effect of dust self-absorption is
negligible, so the Galaxy was considered transparent at these wavelengths.

The linear resolution of the calculations was up to 25\,pc, which is
easily sufficient to model the resolved latitude profiles for structures at the Galactic Centre. In addition the data, which was highly resolved, showed no additional structure (e.g. a thinner layer in z) with sizes below the resolution of the code. For the optimization of the infrared radiation fields, the relevant angular resolution is that of IRAS and Planck bands, which is $5^{\prime}$, corresponding to a linear resolution of approximately 12\,pc at the Galactic Centre. The equivalent numbers for COBE (tracing direct stellar light from old stars in the NIR/MIR) is around $40^{\prime} /90$\,pc.
The optical constants (from UV to submm) of the dust model used in the
computations were those of Weingartner \& Draine (2001) and Draine \& Li
(2007), whose grain model incorporated a mixture of silicates,
graphites and PAH molecules. These optical constants are appropriate
to model diffuse interstellar dust in the MW, as Draine \& Li (2007)
optimized the relative abundances and grain size distributions of the
chemical constituents to fit the extinction law and IR/submm
emissivity of translucent high latitude Cirrus dust clouds in the
solar neighbourhood. The model for the dust emission incorporates a
full calculation of the stochastic heating of small grains and PAH
molecules. As described in PT11, our model accounts for possible
variations in the IR/submm emissivities of grains in dense opaque
molecular dust clouds by employing dust emission templates empirically
calibrated on observed IR/submm emission spectra when accounting for
the emission from such structures.

\section{Fitting the surface brightness photometry from NIR to submm}
\label{sec:fitting}

Fitting the detailed surface brightness photometry of the Milky Way in all
accessible wavelengths is equivalent with optimising for the detailed
geometry and amplitude (luminosity/opacity) of the stellar populations and of the dust. Taking into
account the many geometrical components needed to fit our Galaxy, searching the
whole parameter space with full radiative transfer calculations is 
computationally
prohibited. Therefore we had to develop an intelligent searching
algorithm, which takes into account the orthogonality of the
parameters and avoids degeneracies without involving unnecessary
combinations of  parameters. The concept of this algorithm is to make use of the fact that
  different geometrical and amplitude parameters affect preferentially the emission at specific
wavelengths.  In PT11  it was already shown that different global
parameters affect preferentially  the global emission at specific
wavelengths. Here we confirmed this to also be the case for the
parameters describing the surface brightness distributions.

\begin{figure*}
\centering
\includegraphics[scale=0.7, trim = 0 0 0 0cm, clip]{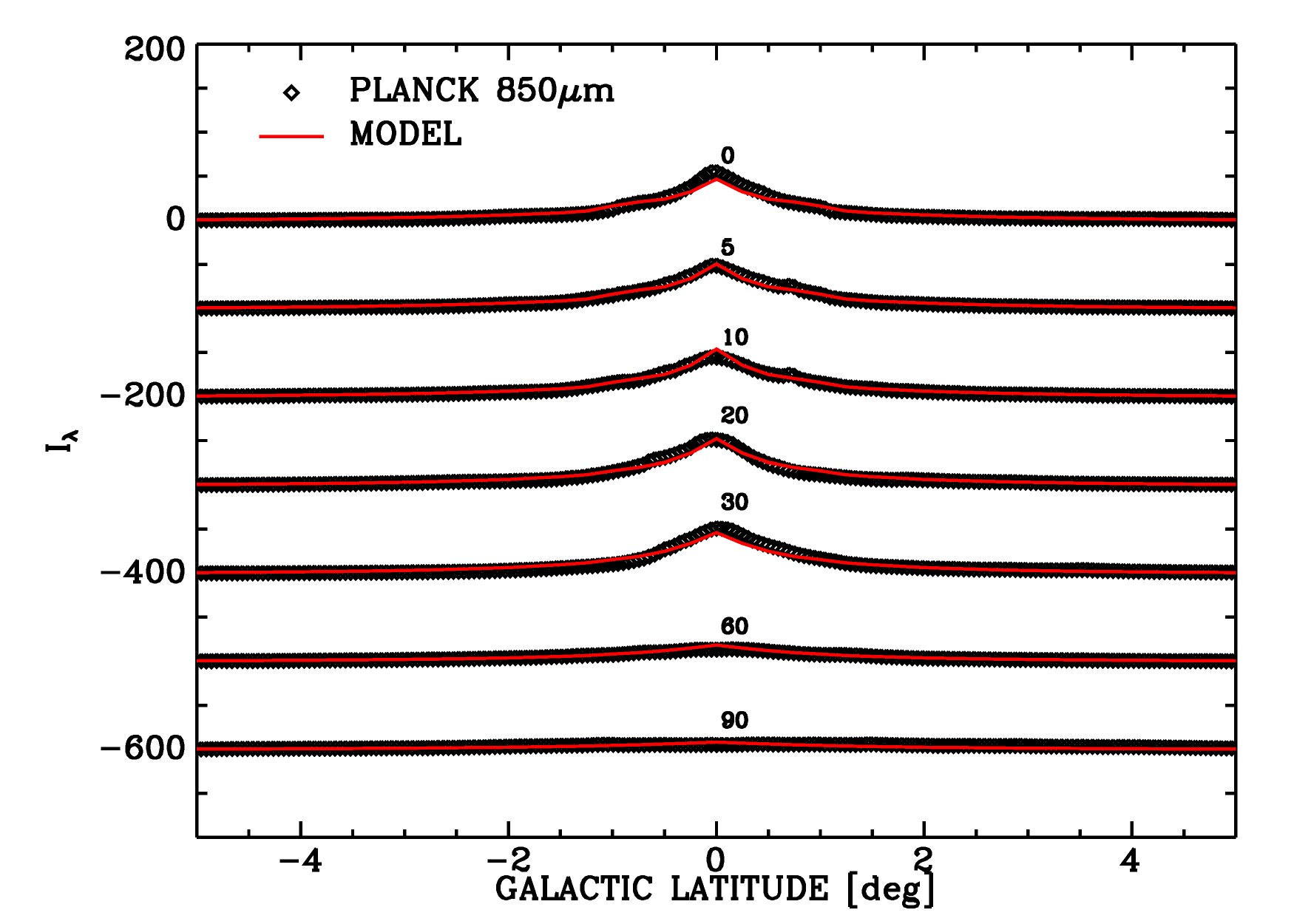}
\caption{Latitude profiles within small longitude bins at 850 $\mu$m. } 
\label{fig:dust_profiles_850um}
\end{figure*}

The first step was to run our generic model from PT11 scaled to some
initial guess of the global parameters of the MW, taken either from
the literature or from general trends of external galaxies. This
allowed us to produce model maps of the MW at all wavelengths, and,
subsequently, 
averaged longitude and latitude profiles (see Figs.~\ref{fig:dust_profiles_1} - \ref{fig:j_profile}), as well as latitude profiles for narrow strips in longitude (see Fig.~\ref{fig:dust_profiles_850um}). The averaged profiles of the model emission
were derived in the same way as those obtained for the
observed images (see Sect.~\ref{sec:data}), allowing thus for a direct
comparison between data and model, which formed the basis for the
optimisation process.
The following steps were taken in this process:\\

\noindent
1. The emission at 25\,${\mu}$m was used to constrain the geometry
of the very young stellar disk and associated dust, represented in our
model by the
clumpy component in form of star-forming clouds. This is because at
this wavelength the emission is dominated by radiation coming from
star-forming regions, where dust is locally heated by the strong radiation
fields of the young massive stars within the birth clouds (Popescu et al. 2002,
Hippelein et al. 2003, Popescu et al. 2005, Sauvage et al. 2005, Hinz et
al. 2006, Popescu et al. 2011). At 25\,${\mu}$m there is also a contribution from stochastically heated dust
grains in the diffuse component, but this is not dominant at this wavelength. We therefore started the
optimisation process by comparing the model profiles with the corresponding 
observed ones at 25\,${\mu}$m (see  
Fig.~\ref{fig:dust_profiles_3}). We first noticed  that the longitude profile is flat for radii
smaller than a characteristic radius, $R_{\rm in}$, rather than increasing 
exponentially towards the centre, as in our generic model. This meant that for 
radii less than $R_{\rm in}$ we had to modify the functional form of the 
emissivity to be a simple linear function, described by the parameter $\chi$
(see Eqn.~\ref{eq:chi}). We found that a linear decrease of emissivity with $\chi=0.5$ produces the 
observed flattening of the emission at 25\,$\mu$m, as seen projected from 
the position of the Sun. The inner radius $R_{\rm in}$ for which the
flattening occurs was unambigously derived  to be 4.5\,kpc. Beyond the inner 
radius the emissivity remains an exponential, like in the generic model, but
with a very abrupt fall-off with increasing radius. Essentially most of the
emission at this wavelength comes from this very compact 
(less radial extended) component. Because
at other infrared wavelengths we found the emission to be more radially
extended, we had to describe the emission at 25\,$\mu$m with a separate
morphological component, which we call inner thin disk. While $R_{\rm in}$ and 
$\chi$ were relatively easily constrained from the $25\,{\mu}$m data, the
scalelength $h_s^{\rm in-tdisk}$ and scaleheigh $z_s^{\rm in-tdisk}$ of the 
inner thin disk had to be constrained by running a grid of models for various 
combinations of these parameters. Strip profiles in latitude helped us to 
constrain a small taper for the scaleheight, with $z_s^{\rm in-tdisk}$ 
increasing linearly with radius. Thus, the optimisation of the $25\,{\mu}$m data allowed 
us to constrain $R_{\rm in}$, $\chi$, $h_s^{\rm in-tdisk}$, and 
$z_s^{\rm in-tdisk}$, as well as the amplitude parameter ${\rm SFR}^{\rm in-tdisk}
\times F^{\rm in-tdisk}$.\\

\noindent
2. Using constraints from the PT11 model, we fixed the scaleheight of
the young stellar disk and of the thin dust disk to be the same as
that of the inner thin disk. Thus we set $z_s^{\rm tdisk}=z_s^{\rm in-tdisk}$, 
$z_d^{\rm tdisk}=z_s^{\rm in-tdisk}$.\\

\noindent
3. The $850\,{\mu}$m Planck band is situated deep in the Rayleigh-Jeans side 
of the emission coming from the diffuse dust component. It is therefore a good
tracer of dust column density. As shown in PT11, the spatially integrated SED 
of spiral galaxies scales mainly with the dust opacity, and is less sensitive 
to the luminosity of the heating sources. It is therefore ideal to constrain 
the distribution of diffuse dust. We thus considered this wavelength for the
3rd step in the optimisation. We ran a new RT calculation with the new values 
of the parameters constrained in steps $1-2$,  and compared the model profiles 
with the corresponding observed ones at $850\,{\mu}$m (see top row in 
Fig.~\ref{fig:dust_profiles_1}). This allowed us to constrain the
parameters of the dust disk and some of the parameters of the thin
dust disk.  As with the $25\,{\mu}$m data, the same
inner flattening of the radial profiles was observed, which was found to be
reproduced by the same linear decrease of the dust opacity within the inner
radius. Thus we were able to constrain the parameters $\chi_d^{\rm disk}=\chi$, 
 $\chi_d^{\rm tdisk}=\chi$, $R_{\rm in,d}^{\rm disk}=R_{\rm in}$ and 
$R_{\rm in,d}^{\rm tdisk}=R_{\rm in}$. Then, we constrained the scalelength 
and scaleheight of the dust disk, $h_{\rm d}^{\rm disk}$ and  $z_{\rm d}^{\rm disk}$, by 
running a grid of models for these parameters. Finally the amplitude
parameters, the opacity of the first and second dust disks, 
$\tau^{\rm f,disk}(B)$ and $\tau^{\rm f,tdisk}(B)$ were derived. The amplitude
parameters had to be readjusted in further steps of the optimisation scheme, in
particular due to changes in the stellar luminosity parameters. Overall, the
optimisation of the  $850\,{\mu}$m data allowed us to constrain 
 $h_{\rm d}^{\rm disk}$, $z_{\rm d}^{\rm disk}$, $\tau^{\rm f,disk}(B)$, $\tau^{\rm
  f,tdisk}(B)$ and set $R_{\rm in}$ and $\chi$ for the dust distributions.
\\

\begin{figure*}
\includegraphics[scale=0.5]{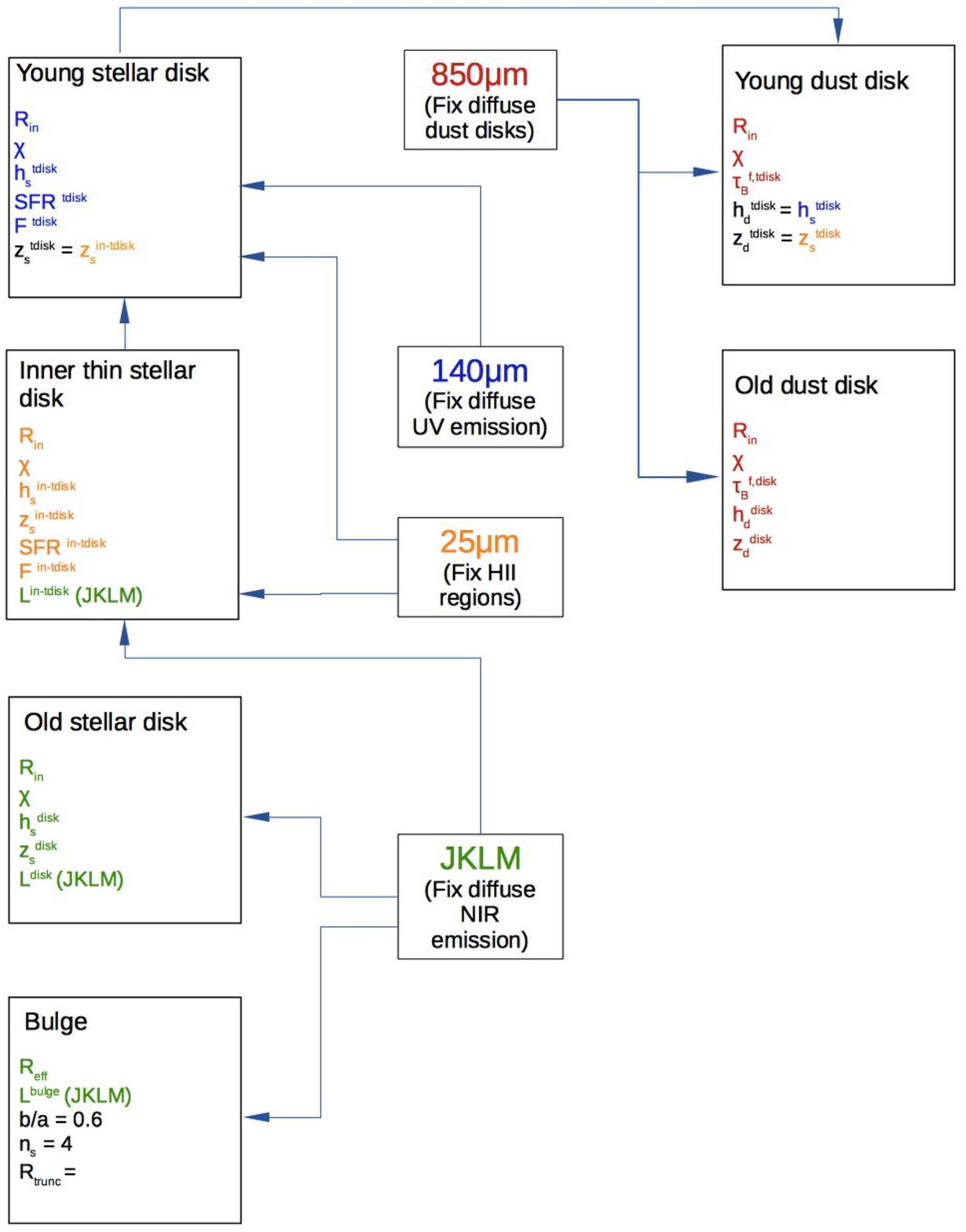}
\caption{A schematic representation of the optimisation algorithm.}
\label{fig:scheme}
\end{figure*}

\noindent
4. Using the constraints from steps $1-3$, we ran a new RT calculation and
compared profiles at the  peak of the dust emission, at $140-240\,{\mu}$m, 
where the emission is strongly influenced by the heating from the young 
stellar disk. Since the young stellar disk cannot be directly
contrained, in the absence of UV observations,  the $140-240\,{\mu}$m
were esential in determining the parameters of the UV emitting disk. 
As with the $25\,{\mu}$m and $850\,{\mu}$m data, the same
inner flattening of the radial profiles was observed, which was found to be
reproduced by the same linear decrease of the stellar emissivity in the thin
stellar disk, within the inner radius. Thus we were able to constrain the 
parameters $\chi_s^{\rm tdisk}=\chi$ and $R_{\rm in,s}^{\rm tdisk}=R_{\rm in}$.
The stepeness of the exponential profile outside $R_{\rm in}$  allowed us to 
constrain the scalelength of the young stellar disk, $h_s^{\rm tdisk}$, and the
overall scaling of the emission constrained the parameter 
${\rm SFR}^{\rm tdisk} \times (1-F)^{\rm tdisk}$. Thus, the
optimisation of the  $140-240\,{\mu}$m data allowed us to constrain 
 $h_s^{\rm tdisk}$, ${\rm SFR}^{\rm tdisk} \times (1-F)^{\rm tdisk}$ and set 
$R_{\rm in}$ and $\chi$ for the stellar emissivity of the young stellar disk.\\

\noindent
5. Using the previous constraints from steps $1-4$, we ran a new RT 
calculation and compared profiles at the NIR wavelengths 
(see Fig.~\ref{fig:dust_profiles_nir}), where we see the 
dust attenuated stellar emission from the old stellar populations. 
We found that the emission has a strong
contribution from a compact component, with the same radial extent as the 
clumpy component visible at $25\,{\mu}$m. We therefore modelled the emission 
with both the standard old stellar disk from PT11, plus the 
inner thin disk. In addition we had to again invoke a linear decrease in the 
stellar emissivity
in the old stellar disk in order to reproduce the flattening of the radial
profiles in the inner regions, with the bulge component superimposed on a 
plateau profile. Thus we adopted the geometrical parameters of the
inner thin disk already fixed at $25\,{\mu}$m, and we optimised for the
scalelength and scaleheight of the old stellar disk,  $h_{\rm s}^{\rm disk}$ and
$z_{\rm s}^{\rm disk}$. The strips latitude profiles allowed us to infer a relatively
strong flare for the scaleheight of the old stellar disk. In the same step we 
also derived the
bulge parameters $R_{\rm eff}$, $b/a$, $L_{\rm J,K,L,M}^{\rm bulge}$ and the amplitude parameters of
the old stellar disk $L_{\rm J,K,L,M}^{\rm
   disk}$, and of the inner thin disk, $L_{\rm J,K,L,M}^{\rm in-tdisk}$. 
Thus, the optimisation of the NIR data allowed us to constrain 
 $h_{\rm s}^{\rm disk}$, $z_{\rm s}^{\rm disk}$, $L_{\rm J,K,L,M}^{\rm
   disk}$, $L_{\rm J,K,L,M}^{\rm in-tdisk}$, $R_{\rm eff}$, 
$b/a$, $L_{\rm J,K,L,M}^{\rm bulge}$, and set $R_{\rm in}$ and $\chi$ for the stellar emissivity of 
the old stellar disk.\\

\noindent
6. Using the previous constraints we ran a new RT calculation and
compared plots at all available wavelengths. Various rescalling of the
global parameters $\tau^{\rm f,disk}(B)$, $\tau^{\rm f,tdisk}(B)$, ${\rm SFR}^{\rm
  tdisk}$,  ${\rm SFR}^{\rm in-tdisk}$, $L_{\rm J,K,L,M}^{\rm disk}$,  $L_{\rm
  J,K,L,M}^{\rm in-tdisk}$ and $L_{\rm J,K,L,M}^{\rm bulge}$ were needed to produce adequate
fits at all wavelengths. This means that the whole process needed several
iterations in order to converge towards the observed surface brightness 
distributions at all wavelengths. A schematic view of the optimisation
procedure is depicted in Fig.~\ref{fig:scheme}.

Inspection of the profiles from Figs.~\ref{fig:dust_profiles_1} - \ref{fig:dust_profiles_nir} shows an overall good
agreement between model and observations. We did not attempted to fit
the 12 micron data, as this is sensitive to PAH abundance, which in
our model is fixed and not varied. At 1.2 micron (Fig.~\ref{fig:j_profile}) we cannot reproduce
the emission within the 4.5 kpc, but this may be due to the
more complex geometry of the inner galactic region. As mentioned in Sect.~\ref{sec:model}, the Milky Way has 
a peanut/boxy bulge/bar within the inner 4.5 kpc, probably dominating the emission at this wavelength, but we have not explicitly included such a component in the model. We only
consider a classical ellipsoidal bulge in the model.

The fitted values of the geometrical parameters of the model (the free parameters) are listed in Table~\ref{tab:geom}. The values of the remaining geometrical parameters (the fixed ones) are given in Table~\ref{tab:geom_fix}.
The derivation of the uncertainties in the main geometrical and amplitude parameters of the model
(those that are constrained from data) is described in 
Appendix~\ref{sec:errors}.

\section{Results}
\label{sec:results}

\subsection{Global properties of the MW}

\subsubsection{The intrinsic SED of the Milky Way}
\label{subsec:intrin_sed}

\begin{figure}
\centering
\includegraphics[scale=0.5]{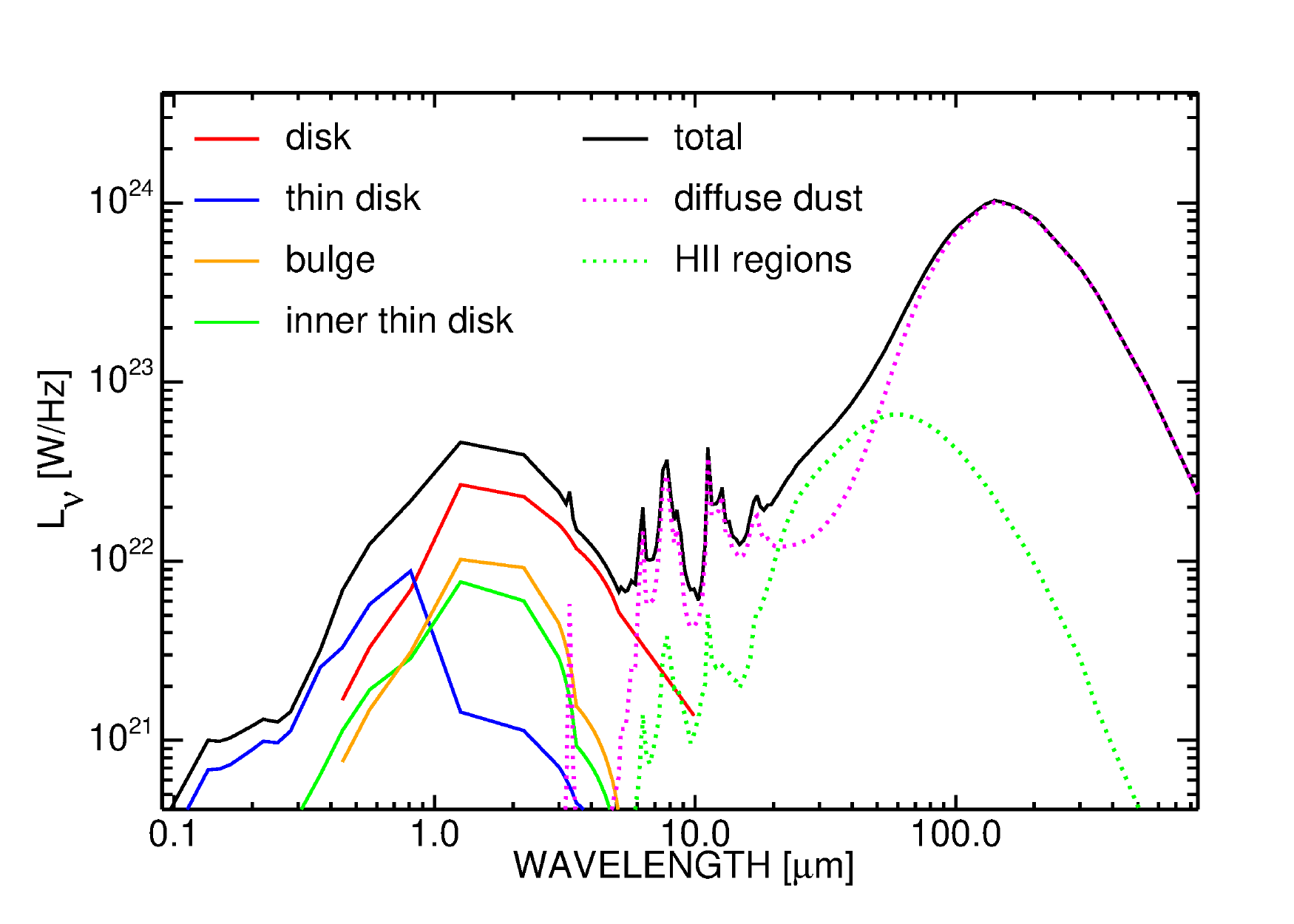}
\caption{Predicted intrinsic (dust corrected) SED of the Milky Way, as would be seen by an observer located outside our Galaxy.} 
\label{fig:intrinsic_SED}
\end{figure}

One of the main results to come out of this work is the
derivation of the intrinsic SED of the Milky Way. This is
shown in Fig.~\ref{fig:intrinsic_SED}, together with the different components
contributing to the global SED. As expected, in the optical
region the emission is dominated by the old stellar disk
and the bulge, while in the UV the emission is dominated
by the thin stellar disk. Almost half ($46\%$) of the stellar luminosity originates from the old stellar disk, with the rest being approximately equally distributed between the young stellar disk and the bulge plus the inner stellar disk. 

In the infrared the emission is
dominated by the diffuse component for wavelengths larger
than 50 µm, and by the clumpy component shortwards of
this wavelength. In the PAH region the emission reverts
to being dominated by the diffuse component. This is in
qualitative agreement with results obtained from other external
galaxies (e.g. NGC891 - Popescu et al. 2011 or M33 -
Thirlwall et al. 2020). Overall the diffuse component of the Milky Way contributes $81\%$ of the total dust emission. We  predict that $(16\pm 1)\%$ of the stellar luminosity is absorbed by dust and re-emitted in the infrared/submm, which is a typical value for early type spirals (Popescu \& Tuffs 2002, Bianchi et al. 2018), but is much smaller than, for example, that 
of M33, for which a value of 
$(35\pm 3) \%$  was derived in Thirlwall et al. (2020), using the same type of models. This is in agreement with the fact that M33 reaches a higher surface density of SFR, as we will discuss in Sect.~\ref{subsec:ssfr}. This shows that the Milky Way is more quiescent, in agreement with its UV/optical colours being redder
than those of NGC891 (Popescu et al. 2011) and M33 (Thirlwall et al. 2020). 

The energy balance between dust absorption and re-emission is found to be dominated by  the young stellar populations. Thus 
$71\%$ of the dust luminosity of the Milky Way is predicted to be powered by the young stars in the thin stellar disk and inner thin disk. This fractional contribution ($F^{\rm dust}_{\rm young}$) is similar to that derived for NGC891 ($69\%$) in Popescu et al. (2011), but somewhat smaller than that derived for M33 ($80\%$) in Thirlwall et al. (2020). These fractions are systematically larger than those derived by Nersesian et al. (2019) for galaxies of similar Hubble type, although a direct comparison is difficult for two reasons. Firstly, the models of Nersesian et al. (2019) are not based on radiative transfer calculations, but only on overall energy balance methods. 
Secondly, our definition of \lq\lq young\rq\rq\ and \lq\lq old\rq\rq\ is in terms of geometrical components rather than stellar age. Thus we call \lq\lq young\rq\rq\ all the stars within the thin (vertical scale-height ranging between 50-90\,pc) disk components, and \lq\lq old\rq\rq\ all the stars in the disk and bulge. On the other hand other radiative transfer studies of galaxies found the young stellar populations to dominate the dust heating, although with a large spread ($63\%$ for M51 in de Looze et al. 2014, between 60 to 80$\%$ for M33 in Williams et al. 2019, $50.2\%$ for M81 in Verstocken et al. 2020, $83\%$ for NGC1068 in Viaene et al. 2020,  $\sim 59\%$ for a sample of 4 barred galaxies in Nersesian et al. (2020a) and $71.2\%$ for M51 in Nersesian et al. 2020b).

\begin{figure}
\centering
\includegraphics[scale=0.31]{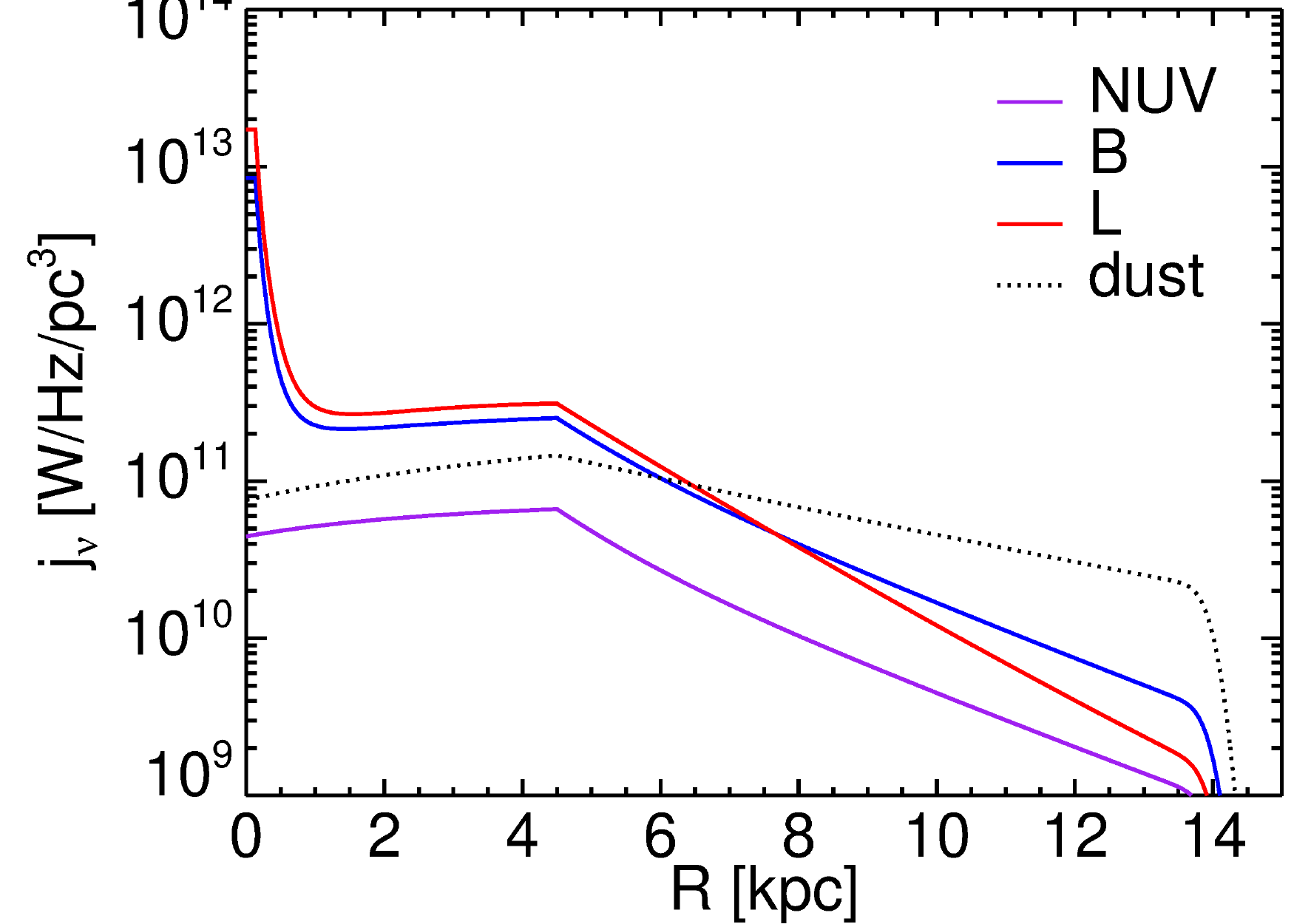}
\includegraphics[scale=0.31]{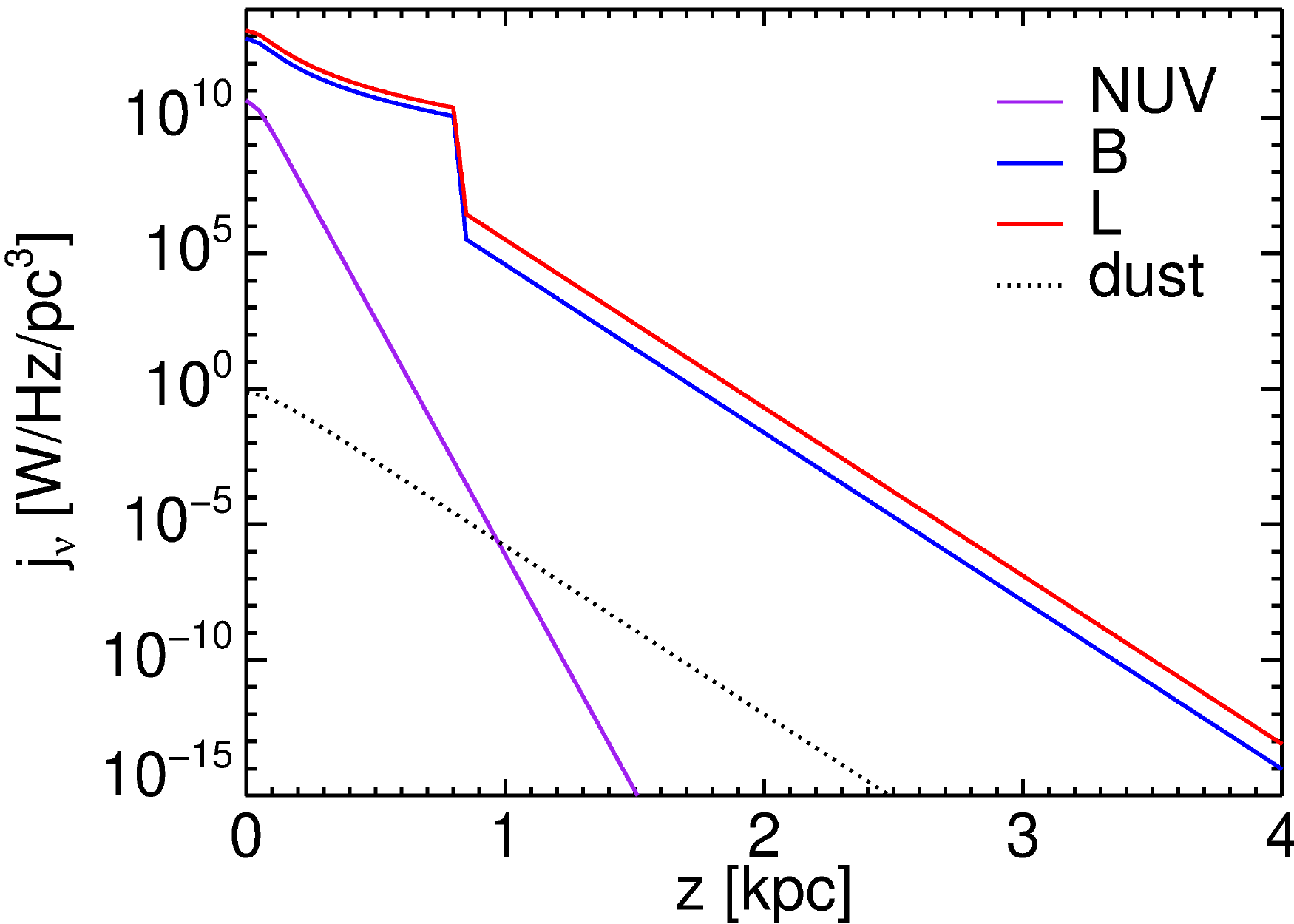}
\caption{Intrinsic stellar emissivity distributions at selected UV/optical wavebands. Top: radial profiles at $z=0$. Bottom: vertical profiles at $R=0$. The profiles of dust opacity (arbitrary scaled) are overplotted as black dotted lines.} 
\label{fig:profiles}
\end{figure}

\subsubsection{Star-formation rate}
\label{sec:sfr}

The star-formation rate of the Milky Way is an important quantity, not only for
the understanding of the formation history of our Galaxy, but also as a
calibrator for external galaxies. Yet, there has been a huge scatter in the
various estimates provided by the different methods employed, ranging from
$0.5-10\,{\rm M}_{\odot}$/yr. The past methods involved
different techniques, like ionization rates derived from radio free-free
emission (Smith et al. 1978, G\"usten \& Mezger 1982, Mezger 1987), from NII
205\,$\mu$m line emission (Bennett et al. 1994, McKee \& Williams (1997), or from WMAP
free-free emission (Murray \& Rahman 2010), SN rates derived
from O/B star counts (Reed 2005), nucleosynthesis measurements derived
from gamma-ray data (Diehl et al. 2006) and YSO star counts
(Robitaille \& Whitney 2010, Davies et al. 2011). The SFR derived in
this paper is obtained by using,  for the first time,  far-infrared data at the peak
of the dust emission SED as the main
constraint, in combination with a radiative transfer
method to link it to the emission from the recently formed
stars. Because of this our approach is complementary to the previous methods. 

The SFR in our model is derived from the intrinsic luminosity of the diffuse and clumpy component in the thin stellar disk and inner thin stellar disk. As described in Sect.~\ref{sec:thin_disk}, the conversion between luminosity and SFR is given by Eqs.~16-18 from PT11.
We derive a ${\rm SFR}=1.25\pm0.2\,{\rm M}_{\odot}$/yr, which is in the 
range of values with the most recent determinations of $1.9\pm0.4\,{\rm
  M}_{\odot}$/yr by Chomiuk \& Povich (2011) and  $1.65\pm0.19\,{\rm
  M}_{\odot}$/yr by Licquia \& Newman (2015). Most of the obscured star
formation occurs in our model in the inner 4.5 kpc, within the inner thin stellar
disk, with ${\rm SFR}^{\rm in-tdisk}=0.25\,{\rm M}_{\odot}$/yr. The rest of $1.0\,{\rm
  M}_{\odot}$/yr is distributed in the thin stellar disk, with most of
the UV photons escaping and powering the diffuse component ($F^{\rm
  tdisk}=0$).

\subsubsection{Stellar mass}
\label{sec:stellar_mass}

We got a simple estimate of the stellar mass $M_*$ using 
mass-to-luminosity ratios calibrated in terms of 
color-magnitude diagrams for external galaxies. For this we 
use the optical calibration from Taylor et al. (2011). By applying this calibration we 
obtain a stellar mass of $M_*=(4.9\pm 0.3)\times10^{10}\,{\rm M}_{\odot}$, which agrees
quite well with the other estimates from literature. Thus in the 
review of Bland-Hawthorne \& Gerhard a dynamic stellar mass 
of the MW of $(5\times 10^{10})\pm(1\times10^{10})\,{\rm M}_{\odot}$ is quoted, while the photometric derived stellar 
mass of Flynn et al. (2006) is in the range $(4.85-5.5)\times 10^{10}\,{\rm M}_{\odot}.$\\\\

\subsubsection
{Specific star-formation rate and surface density}
\label{subsec:ssfr}

Using the derived SFR from Sect.~\ref{sec:sfr} and the stellar mass derived in Sect.~\ref{sec:stellar_mass} we calculate a specific star-formation rate sSFR, defined as the star formation rate per unit stellar mass, to be ${\rm sSFR}=2.6\pm 0.4 \times 10^{-11}$\, yr$^{-1}$. This is similar to the value of $(2.71\pm 0.59)\times 10^{-11}$yr$^{-1}$ derived by Licquia \& Newman (2015) using Bayesian methods to analyse various measurements from the literature. 

We also derive a surface density of star-formation rate, $\Sigma_{\rm SFR}$, by using the area of the disk out to the truncation radius of the model, $R_{\rm t}=14\,$kpc. We obtain $\Sigma_{\rm SFR}=(2\pm 0.3)\times 10^{-3}{\rm M}_{\odot}\,{\rm yr}^{-1}\,{\rm kpc}^{-2}$. The obscured SFR has a higher surface density, with $\Sigma^{\rm obsc}_{\rm SFR}=3.9\times 10^{-3}{\rm M}_{\odot}\,{\rm yr}^{-1}\,{\rm kpc}^{-2}$ within the inner 4.5\,kpc. These numbers point towards MW being a relatively quiescent galaxy, as already anticipated in  Sect.~\ref{subsec:intrin_sed}. For example the nearby M33 is more active in forming stars,  in particular in the inner region, reaching a higher surface density of SFR, with $\Sigma_{\rm SFR}^{\rm n}=103\times 10^{-3}\, {\rm M}_{\odot}\,{\rm yr}^{-1}$ for the nuclear region, $\Sigma_{\rm SFR}^{\rm n}=10\times 10^{-3}\, {\rm M}_{\odot}\,{\rm yr}^{-1}$ for the inner disk, and $\Sigma_{\rm SFR}^{\rm n}=3\times 10^{-3}\, {\rm M}_{\odot}\,{\rm yr}^{-1}$ for the main disk.

\subsubsection
{Dust mass and dust opacity}

We derived a dust mass for the Milky Way of $4.78\pm0.06 \times 10^{7}\,{\rm M}_{\odot}$. Misiriotis et
al. (2006) derived a mass of dust of $7.02\times10^7\,{\rm M}_{\odot}$, which is higher than our
value. We believe that the main reason for this discrepancy is that in
Misiriotis et al. they did not restrict the modelling to a narrow strip in 
latitude, as we did, and therefore their analysis may be subject 
to contamination from 
higher latitude emission local to the Sun. Looking at gas measurements, the 
COBE non-RT analysis incorporating gas of Sodroski et al. (1997)
estimates $3.5\times10^9\,{\rm M}_{\odot}$ for the HI, and $1.3\times 10^9\,{\rm
  M}_{\odot}$ for the ${\rm H}_2$, which 
means a total gas mass of $4.8\times 10^{9}\,{\rm M}_{\odot}$. This would be in agreement with our 
dust masses for a gas-to-dust ratio of 100, somewhat less than
inferred at the solar circle, but nevertheless reasonable when one
considers the metallicity gradient in the Milky Way, which might be expected
to give rise to a increasing gas-to-dust ratio with increasing Galactocentric
radius.  

The dust opacity has a maximum value at the position of the inner
radius, with $\tau_{\rm B}^{\rm f}(R_{\rm in})=1.48\pm 0.1$. The opacity is
dominated by the main dust disk, with $\tau^{\rm f,disk}/\tau^{\rm f,tdisk}=5.2$.

\subsection{Spatial distributions}
\label{sec:spatial_distrib}

Examples of resulting stellar and dust distributions are plotted in Fig.~\ref{fig:profiles}. The top panel of the figure shows radial profiles at mid-plane ($z=0$) while the bottom panel shows vertical profiles at the centre ($R=0$). The examples are at three selected wavelengths: in the ultraviolet (GALEX NUV), in the optical (B-band) and in the NIR (L band). The dust distributions are also overplotted as black dotted lines. The plots show the overall characteristics of the main constituents of our model: the old stellar disk, the bulge, the young stellar disk, the inner stellar disk and the dust disk. Below we describe the results obtain for their corresponding distributions.

\subsubsection{The old stellar disk}

Knowledge of the scale-length of the old stellar disk of the Milky Way
has been very uncertain, with values in the literature ranging from
1.8 to 6.0 kpc. Since optical estimates are prone to strong extinction,
infrared determinations were instead used to constrain scalelengths (Kent et
al. 1991, Ruphy et al. 1996, Freudenreich 1998, Drimmel \& Spergel
2001, Lopez-Corredoira et al. 2002, Cabrera-Lavers et al. 2005, Benjamin et al. 2005, Reyle
et al. 2009). Bland-Hawthorne \& Gerhard
(2016) analysed existing determinations and produced an average value of
$h_{\rm s}^{\rm disk}=2.6\pm 0.5$\,kpc. This is consistent with our determination
of the scalelength in the K band of $h_{\rm s}^{\rm disk}(K)=2.2\pm 0.6$\,kpc. However,
in our model we allow for a wavelength dependent scale-length, such
that this increases monotonically with decreasing optical
wavelength. Thus, we derived a B-band scalelength of $h_{\rm
  s}^{\rm disk}(B)=3.2\pm 0.9$\,kpc. This value was fixed in our model to be the same as
that of the thin stellar disk, and was constrained from data at the
peak of the dust emission SED. This value seems to be consistent with
results from Bovy et al. (2012) who suggests that younger stellar
populations may have a longer scale-length of 3 kpc or larger.

Determinations of scale-heights of the old stellar disk were
restricted to the solar neighbourhood and were spanning the range of
$z_{\rm s}^{\rm disk}=220-450$\,pc. The recommended value from the review of 
Bland-Hawthorne \& Gerhard (2016) is  $z_{\rm s}^{\rm disk} =300$\,pc, which is
based on Juric et al. (2008). This derived value is identical to our
determination. 
However, unlike existing studies that
only dealt with the local value, our model derived a scale-height
throughout the volume of the Milky Way. Thus, we found a radial
dependent scaleheight showing a moderate flare, with  $z_{\rm s}^{\rm disk}(0)
=140\pm 20$\,pc in the centre, increasing to  $z_{\rm s}^{\rm disk}(R_{\rm in})
=170\pm 20$\,pc at the inner radius and to $z_{\rm s}^{\rm
  disk}(R_{\odot}) =300\pm 20$\,pc
at the solar radius.

\subsubsection{The thin stellar disk}
\label{sec:thin_stellar_disk}

The scalelength of the thin stellar disk mainly emitting in the UV
was found to be $h_{\rm s}^{\rm tdisk}=3.2\pm0.9$\,kpc. No other determinations of this
quantity exists in the literature. Our own determination is constrained from
the FIR data at the peak of the dust emission. In particular the shape of the 
latitude profile is strongly influenced by the value of $h_{\rm s}^{\rm
  tdisk}$, and this is how this length parameter is derived. For the 
scale-heights we found again a radially dependent value, this time with a linear 
taper, such that $z_{\rm s}^{\rm tdisk}(0)=50\pm 10$\,pc and $z_{\rm s}^{\rm
  tdisk}(R_{\odot})=90\pm 10$\,pc.\\

\subsubsection{The inner stellar disk}

This new stellar component was discovered because of two features seen in the
profiles that could not have been explained using the existing stellar
components. Firstly, the 24 microns revealed an HII component that was mainly
distributed within the inner 4.5 kpc. On the other hand the FIR latitude
profiles required a rather extended young stellar disk in order to provide
enough heating to the large scale distribution of diffuse dust to  match the
observations. It appeared then that, unlike in our standard model for external
galaxies, we had to decouple the obscured star forming disk from the young 
stellar disk, and admit that we have an extra inner stellar disk where most of
the recent star formation occurs. Secondly, when looking at the NIR latitude
profiles, there was excess emission visible towards the inner 4.5 kpc region,
not accounted for by the main stellar disk. This emission was therefore
associated to this new inner component. 

Because the inner thin disk has a very small exponential scalelength outside
the inner radius, it can be said that most of its stellar emissivity decreases
linearly with decreasing radius. The vertical distribution is that of the thin
disk, with $z_{\rm s}^{\rm in-tdisk}(0)=50\pm 10$\,pc and $z_{\rm s}^{\rm
  in-tdisk}(R_{\odot})=90\pm 10$\,pc.
It is possible that our inferred inner thin disk is the axi-symmetric counterpart of the 

so-called ``long bar'', an overdensity of sources at positive longitudes
with a wide longitude extent and a narrow extent along the line of sight. The long bar
 was inferred by Hammersley et al. (2000), Cabrera-Lavers et
al. (2007), (2008) and Benjamin et al. (2005) using UKIDSS and GLIMPSE star
counts.  Wegg et al. (2015) investigated the long
bar using RCG stars and found a total bar half length of $5.0\pm0.2$\,kpc. This
is consistent with the compact extent of our inner thin disk within the inner
4.5 kpc and the sharp truncation beyond this radius. These features  and the local maximum in the overall stellar and dust emissivity found at this position are also consistent with the existence of two logarithmic spiral arms, as proposed by Dobbs \& Burkert (2012).

Perhaps the most important feature of our inner disk, not found in other
studies, is the domination of obscured star formation within this
component. This would suggest a close correspondence between the distribution
of the thin inner disk and that of the CO distribution. Indeed, 
our results are in qualitative agreement with the estimates of the CO
surface density, as shown in Miville-Deschenes et al. (2016). Thus in 
both the distribution of CO and of our clumpy component there is a local peak at
around 4.5 kpc, there is a rapid fall-off beyond 4.5 kpc and there is a
decrease within 4.5 kpc. However, there are some quantitative differences in the rates of fall-off, which can either be due to change in excitation of the CO or
to variations in the escape fraction of UV photons from the HII regions with
radial position.

\subsubsection{The dust disk}

Most of the diffuse dust is in the form of a disk with scalelength $h_{\rm
  d}^{\rm disk}=5.2\pm 0.8$\,kpc and scaleheight $z_{\rm
  d}^{\rm disk}=0.14\pm0.02$\,kpc. The scalelength is about 1.6 larger than that of
the young stellar disk, result that is in agreement with studies of external
galaxies (Xilouris et al. 1999, De Geyter et al. 2014). The disk is
thicker than the young stellar disk, but thinner than the old stellar disk,
again in agreement with other studies of external edge-on galaxies (Xilouris et
al. 1999,  De Geyter, G. et al. 2014, 2015). 

The derived distribution of dust can be compared to the HI distribution. The 
atomic hydrogen of the Milky Way was modelled with two disk components, one
with scalelength of 3.75\,kpc and one with scalelength of 7.5\,kpc 
(Kalberla \& Kerp 2009). This is in qualitative agreement with our derived dust
distribution, although a quantitative comparison would depend on the relative 
abundance of grains in these two HI components.

\section{Discussion}
\label{sec:discussion}

\subsection{The role of the different stellar populations in heating the dust}

\begin{figure}
\centering
\includegraphics[scale=0.3]{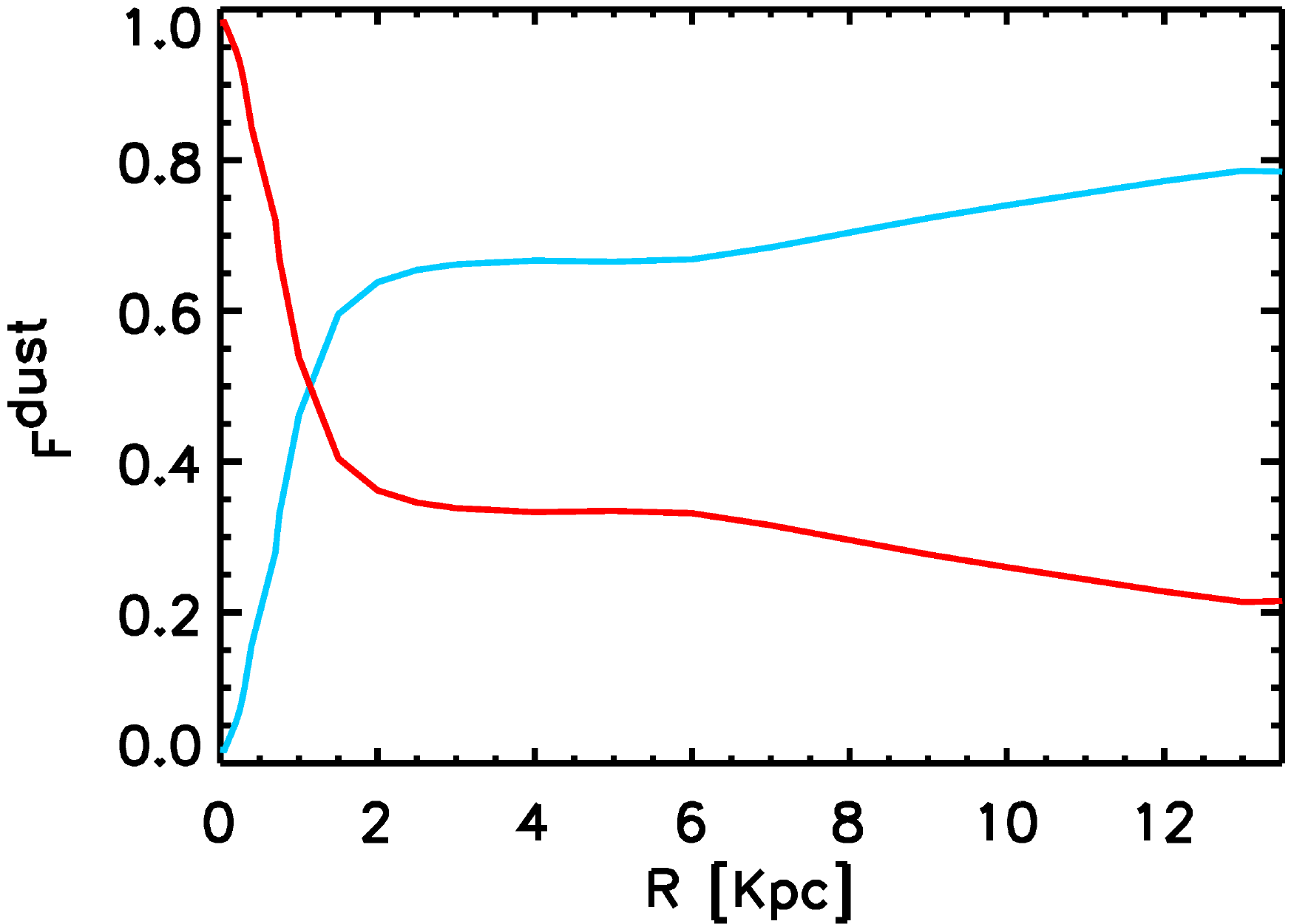}
\caption{Radial profiles of the fraction of stellar light from young (blue line) and old (red line) stellar populations in heating the dust, $F^{\rm dust}_{\rm young}$ and $F^{\rm dust}_{\rm old}$, respectively. They were calculated by integrating the energy absorbed from the different stellar populations over the vertical positions. Here we note that the signature of the Central Molecular Zone was masked out from the data (see Sect.~\ref{sec:data}) and not taken into account in our modelling. Its inclusion would have probably produced a spike  in the $F^{\rm dust}_{\rm young}$ within the inner 200\,pc.}
\label{fig:Eabs_frac}
\end{figure}

In Sect.~\ref{subsec:intrin_sed} we found that 71\% of the total dust luminosity in the Milky Way is powered by the young stellar populations ($F^{\rm dust}_{\rm young}=0.71$) . Although the young stars dominate the heating mechanisms when integrating over the whole Galaxy, this is not always the case when looking at local scales. In Fig.~\ref{fig:Eabs_frac} we show radial profiles of these fractions. They were calculated by integrating the energy absorbed (from the different stellar populations) over the vertical positions for each radial bin. One can see that in the inner 1\,kpc it is the old stellar population that dominates the dust heating. This is due to the strong contribution of the bulge within $\sim 3 R_{\rm eff}$, with $F^{\rm dust}_{\rm old}$ following the decrease in the bulge stellar emissivity with increasing radial distance. At around 1 kpc from the centre there is roughly equal contribution to the dust heating from both old and young stars. Between 1 and 2 kpc $F^{\rm dust}_{\rm old}$ continues to decrease, such that at 2 kpc the young stellar populations become the dominant heating source, with $F^{\rm dust}_{\rm young}(2\,{\rm kpc})\simeq 0.65$. Between 2 and 6 kpc $F^{\rm dust}_{\rm young}$ remains fairly constant, although it is the inner thin disk that dominates the heating until 4.5 kpc, and the thin stellar disk from 4.5 kpc outwards. Between 6 and 14 kpc $F^{\rm dust}_{\rm young}$ shows a shallow monotonic increase, from $F^{\rm dust}_{\rm young}(6\, {\rm kpc})\simeq 0.65$ to  $F^{\rm dust}_{\rm young}(14\, {\rm kpc})\simeq 0.8$. This is mainly due to the decrease in the dust opacity, with UV photons being more readily absorbed than the long wavelength photons. At the solar position $F^{\rm dust}_{\rm young}(R_{\odot})\simeq 0.7$.

\begin{figure}
\centering
\centering
\includegraphics[width=1.0\linewidth]{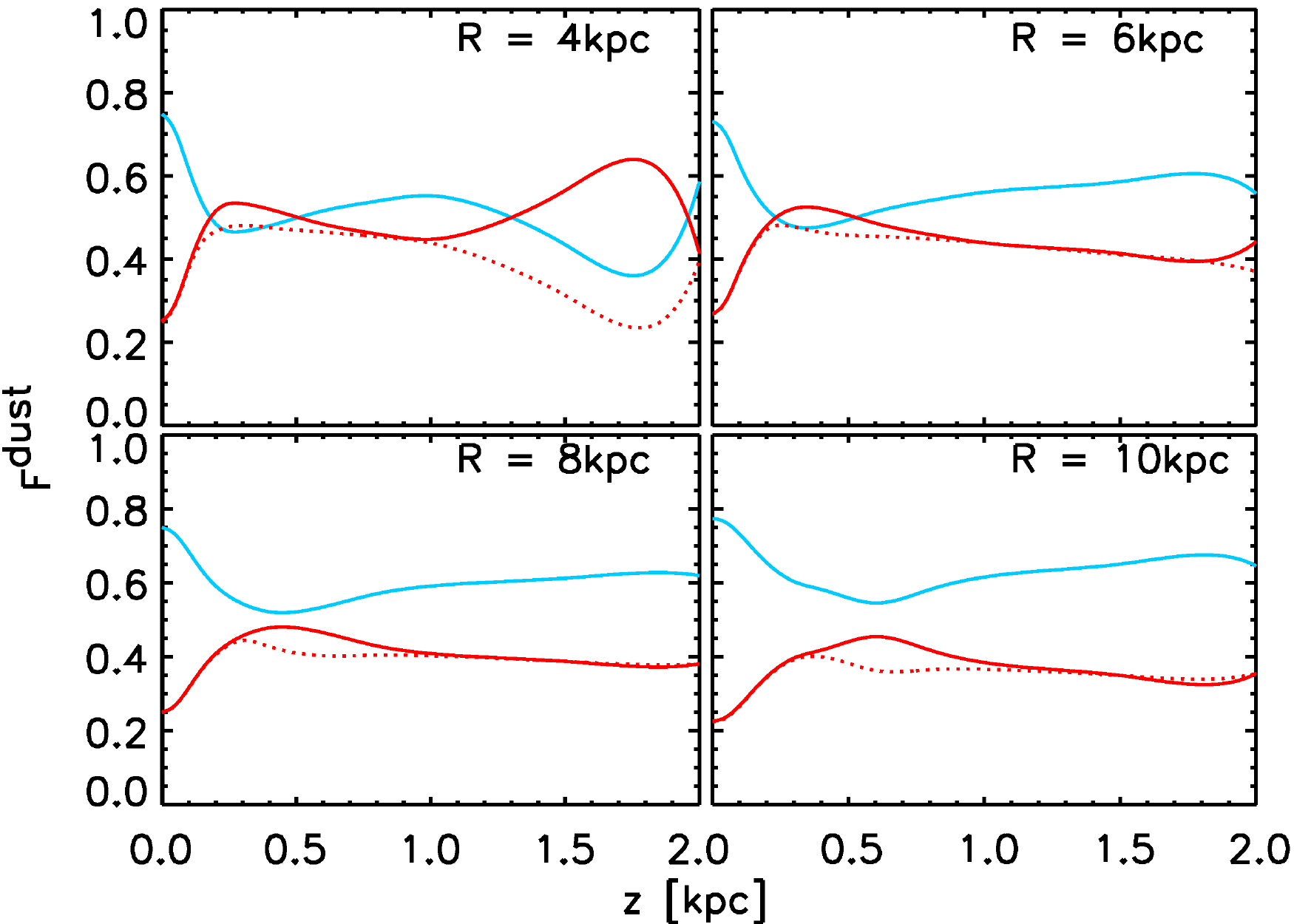}
\caption{Vertical profiles of the fraction of stellar light from young (blue line) and old (red line) stellar populations in heating the dust, $F^{\rm dust}_{\rm young}$ and $F^{\rm dust}_{\rm old}$ respectively,  at $R=4$\,kpc (top left), $R=6$\,kpc (top right), $R=8$\,kpc (bottom left) and $R=10$\,kpc (bottom right).}
\label{fig:Eabs_frac_z}
\end{figure}

The variation of $F^{\rm dust}$ with vertical position is strongly affected by the disk scale-heights and their variation  with radial distance. To illustrate this point we plotted in Fig.~\ref{fig:Eabs_frac_z}  vertical profiles of $F^{\rm dust}$ at $R=4,6,8,10$\,kpc. Here we note that, unlike the radial profiles from Fig.~\ref{fig:Eabs_frac} that are averaged over vertical positions, the vertical plots are strips at a given radius (rather than an average over all radii). The same blue and red lines are used to represent the contribution of the young and old stellar populations. To this we overplotted with dotted red lines the profile of the fraction $F^{\rm dust}_{\rm old, disk}$ (without bulge included).

We remind the reader that the trends shown in the vertical plots from Fig.~\ref{fig:Eabs_frac_z} could be further influenced by a possible halo contribution (not included in our model). Collisionaly heated grains (Gail \& Sedlmayr 1975, Draine \& Anderson 1985, Dwek 1986, Dwek et al. 1990, Dwek \& Arendt 1992, Popescu et al. 2000b, Bocchio et al. 2016) may also play a role high above the disk , although there are no observational constraints for this.

\subsection{Spatial variation of SFR and stellar mass} 

\begin{figure}
\centering
\includegraphics[scale=0.24]{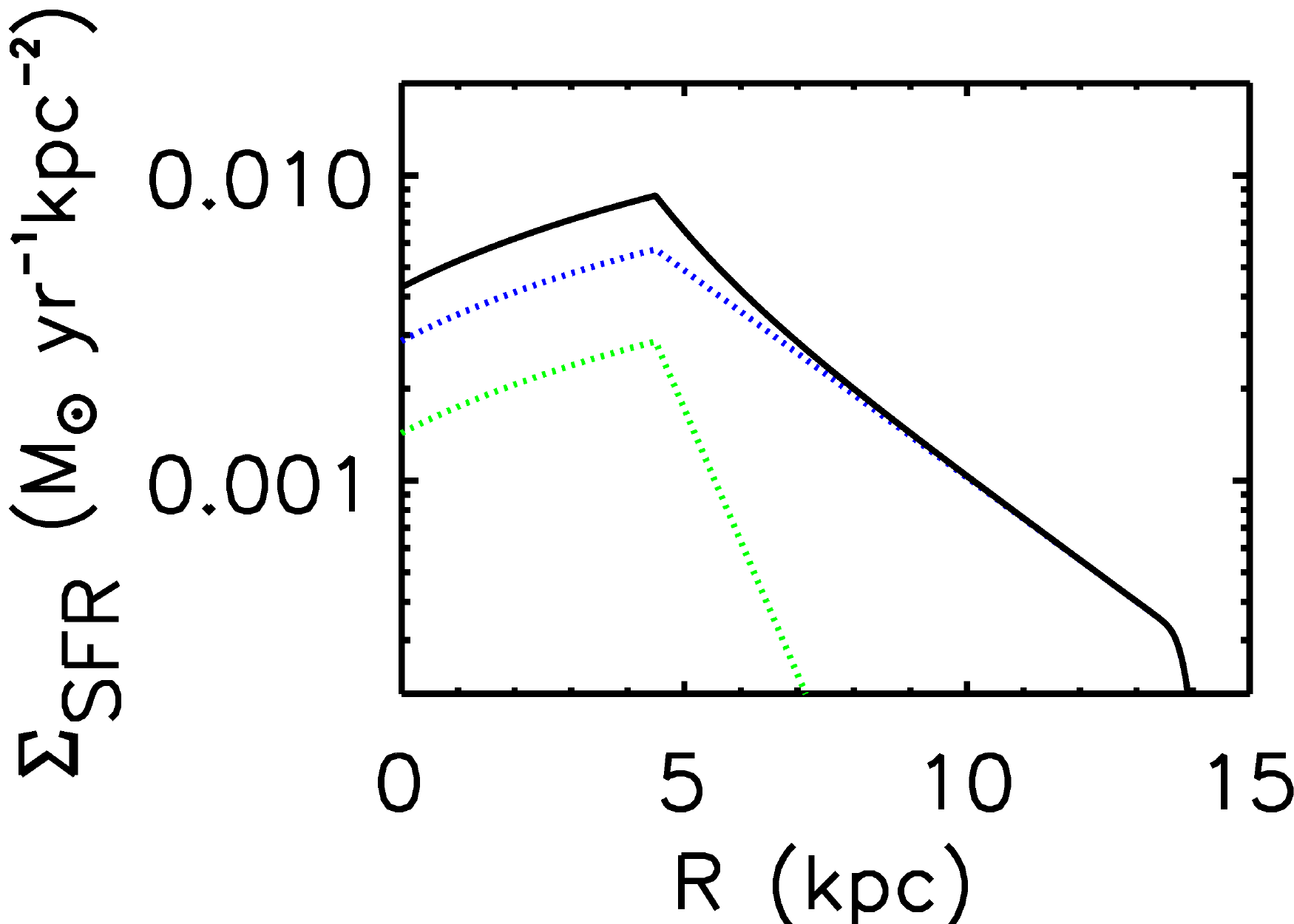}
\includegraphics[scale=0.23]{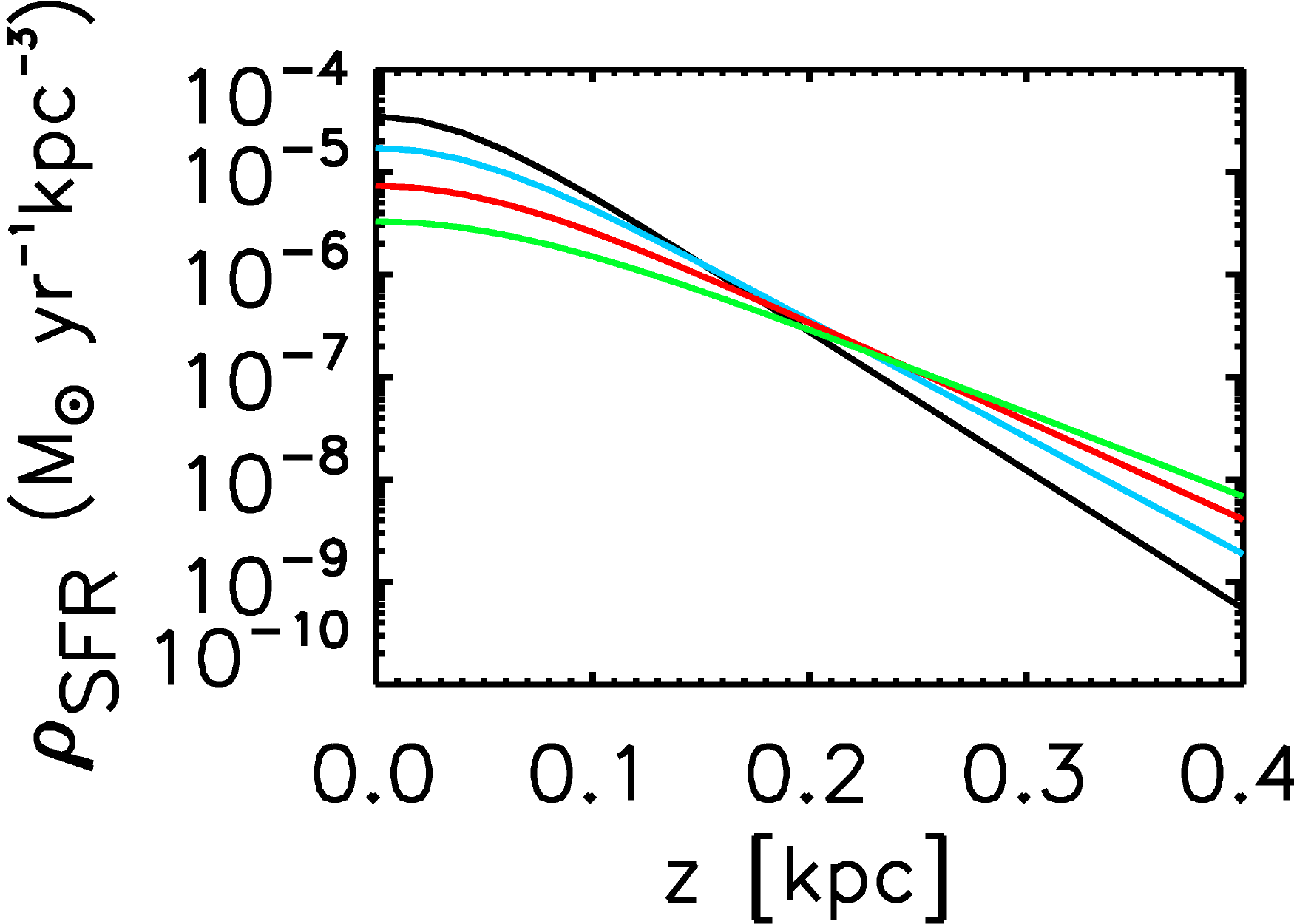}
\caption{Left: Radial profile of star-formation rate surface density,  $\Sigma_{\rm SFR}$ (solid black line). The contribution from the thin disk and thin inner disk to the  $\Sigma_{\rm SFR}$ are shown with blue and green lines, respectively. Right: Vertical profiles of star-formation rate volume density, $\rho_{\rm SFR}$ at different radii are plotted with black ($R=4$\,kpc), blue ($R=6$\,kpc), red ($R=8$\,kpc) and green ($R=10$\,kpc).}
\label{fig:sfr_dens}
\end{figure}

In Sect.~\ref{subsec:ssfr} we found that $\Sigma_{\rm SFR}=(2\pm0.3)\times10^{-3} {\rm M}_{\odot} {\rm yr}^{-1}\, {\rm kpc}^{-2}$ when averaging over the whole of the Milky Way.  $\Sigma_{\rm SFR}$ varies though by two orders of magnitude when moving from the centre to the outer disk. As shown in the left panel of Fig.~\ref{fig:sfr_dens}, there is a maximum of $\Sigma_{\rm SFR}\simeq1\times10^{-2} {\rm M}_{\odot} {\rm yr}^{-1}\, {\rm kpc}^{-2}$ at around $R=4.5$\,kpc, where the inner thin disk (extended bar) ends (has a sharp decline in emissivity). $\Sigma_{\rm SFR}$ falls to as low as 
$\Sigma_{\rm SFR}\simeq 4\times10^{-4} {\rm M}_{\odot} {\rm yr}^{-1}\, {\rm kpc}^{-2}$ at around $R=14$\,kpc, being a factor 10 higher in the centre.

In the right panel of Fig.~\ref{fig:sfr_dens} we show vertical profiles of SFR volume density, $\rho_{\rm SFR}$, for different radii. There is a general decrease of $\rho_{\rm SFR}$ with vertical distance. The decrease is steeper for smaller radii and shallower at larger radii, reflecting the linear increase in scale-height with radius for the young stellar populations. 

\begin{figure}
\centering
\includegraphics[scale=0.23]{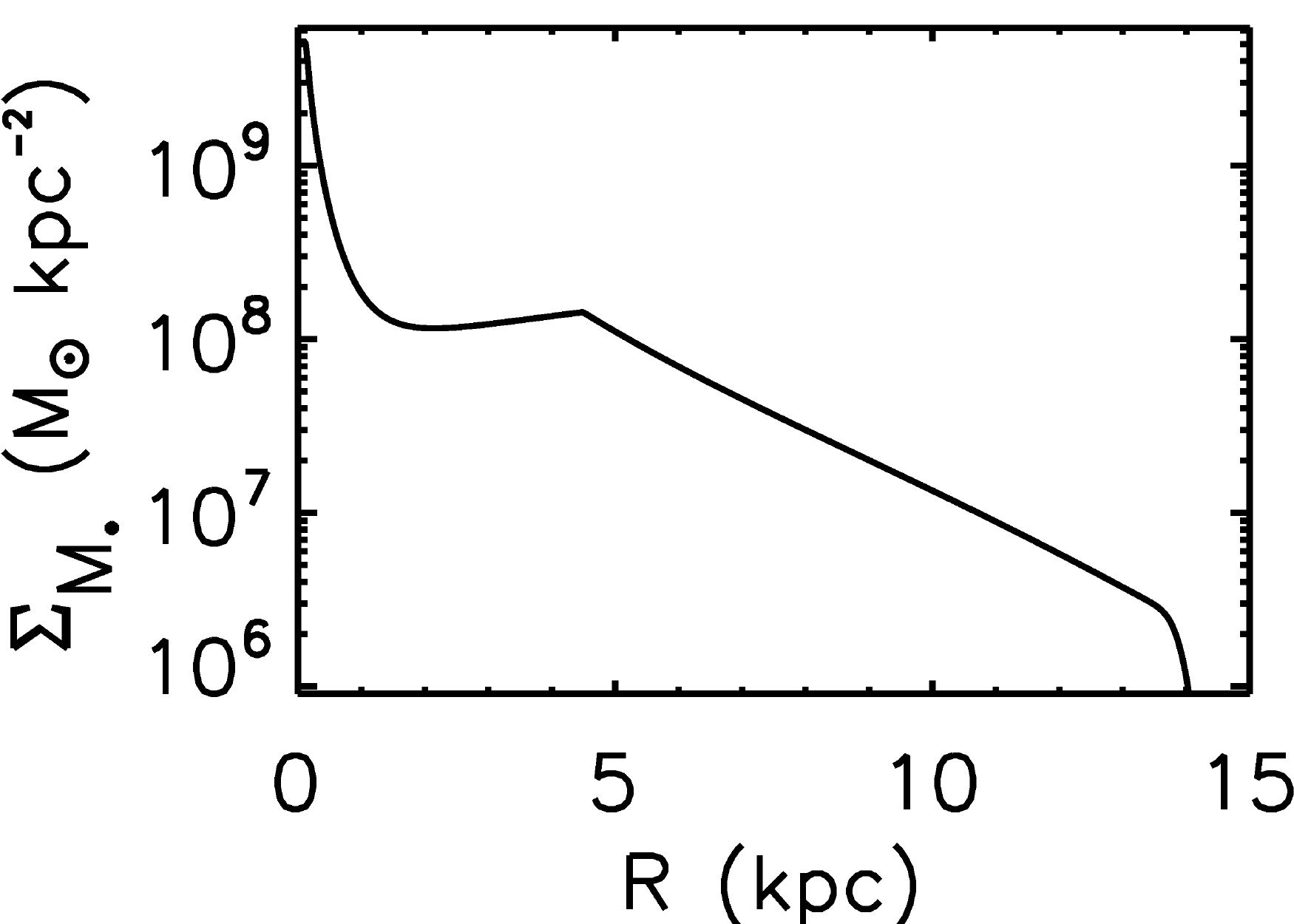}
\includegraphics[scale=0.23]{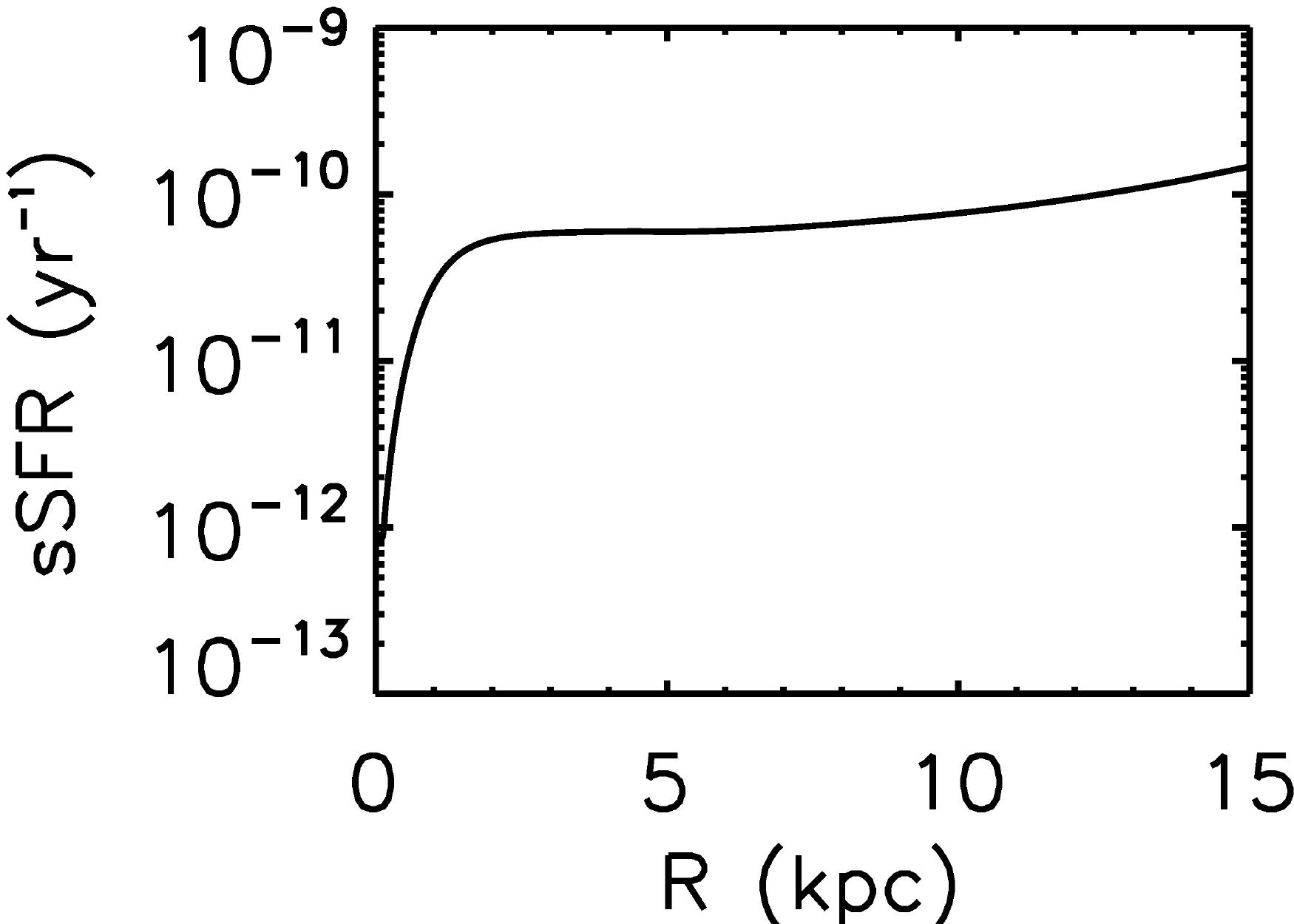}
\caption{Radial profiles of stellar mass surface density $\Sigma_{M_{*}}$ (left) and specific star-formation rate sSFR (right).}
\label{fig:stellar_mass_ssfr}
\end{figure}
The stellar mass surface density, $\Sigma_{M_{*}}$, was derived by scaling the surface density of L-band luminosity to the known value of $\Sigma_{M_{*}}=30\,{\rm M}_{\odot}{\rm pc}^{-2}$ at the solar position. A radial profile is given in the left panel of Fig.~\ref{fig:stellar_mass_ssfr}. The integration of this profile provides a total mass of $M_{*}=(4.6\pm 0.3)\times 10^{10}\,{\rm M}_{\odot}$, which is consistent with our global determination of $M_{*}=4.9\pm 0.3\times 10^{10}\,{\rm M}_{\odot}$ obtained in Sect.~\ref{sec:stellar_mass} using mass-to-luminosity ratios calibrated in terms of color-magnitude diagrams and the optical calibration from Taylor et al. (2011).

The sSFR (right panel of Fig.~\ref{fig:stellar_mass_ssfr}) is relatively constant with radius, except for the inner 1\,kpc, where there is a strong dip, due to the bulge contribution to the stellar mass. However, we caution the reader that the central dip may be overestimated because of  our assumption that the bulge has a simple de Vaucouleurs profile, while in reality  the bulge of the Milky Way has a complex peanut/boxy shape.  In the outer disk, beyond $R=10$\, kpc, sSFR has a mild increase, with no dramatic variation in the slope of the profile. It is therefore reasonable to conclude that the Milky Way has a relatively constant sSFR throughout most of its radial extent (at around ${\rm sSFR}\simeq 5.\times 10^{-11}{\rm yr}^{-1}$).

\subsection{The attenuation curve of the Milky Way}

The average extinction curve of the Milky Way has been used as a standard means for characterising the effects of dust on the observed stellar light 
passing not only through the interstellar medium (ISM) of the Milky Way,
but in general, through within spiral galaxies in the nearby Universe and beyond, since there are no equivalent extinction measurements for galaxies other than the Milky Way. Moreover, attenuation curves derived for spiral galaxies, usually under the assumption of Milky Way extinction characteristics, have been also compared to the average extinction curve of the Milky Way, since, in the absence of a well-defined standard attenuation curve,  this has been the only practical comparison that allows to disentangle  the effects of geometry from the intrinsic properties related to the optical constants of dust grains. But what is the actual attenuation curve of the Milky Way and how different is it from the extinction curve? What would an observer outside the Milky Way  derive, lets say, if they were to see the Milky Way at an average inclination? And what would they derive if they were to see the Milky Way edge one? Here we predict for the first time this fundamental property of the Milky Way.

\begin{figure}
\centering
\includegraphics[scale=0.45]{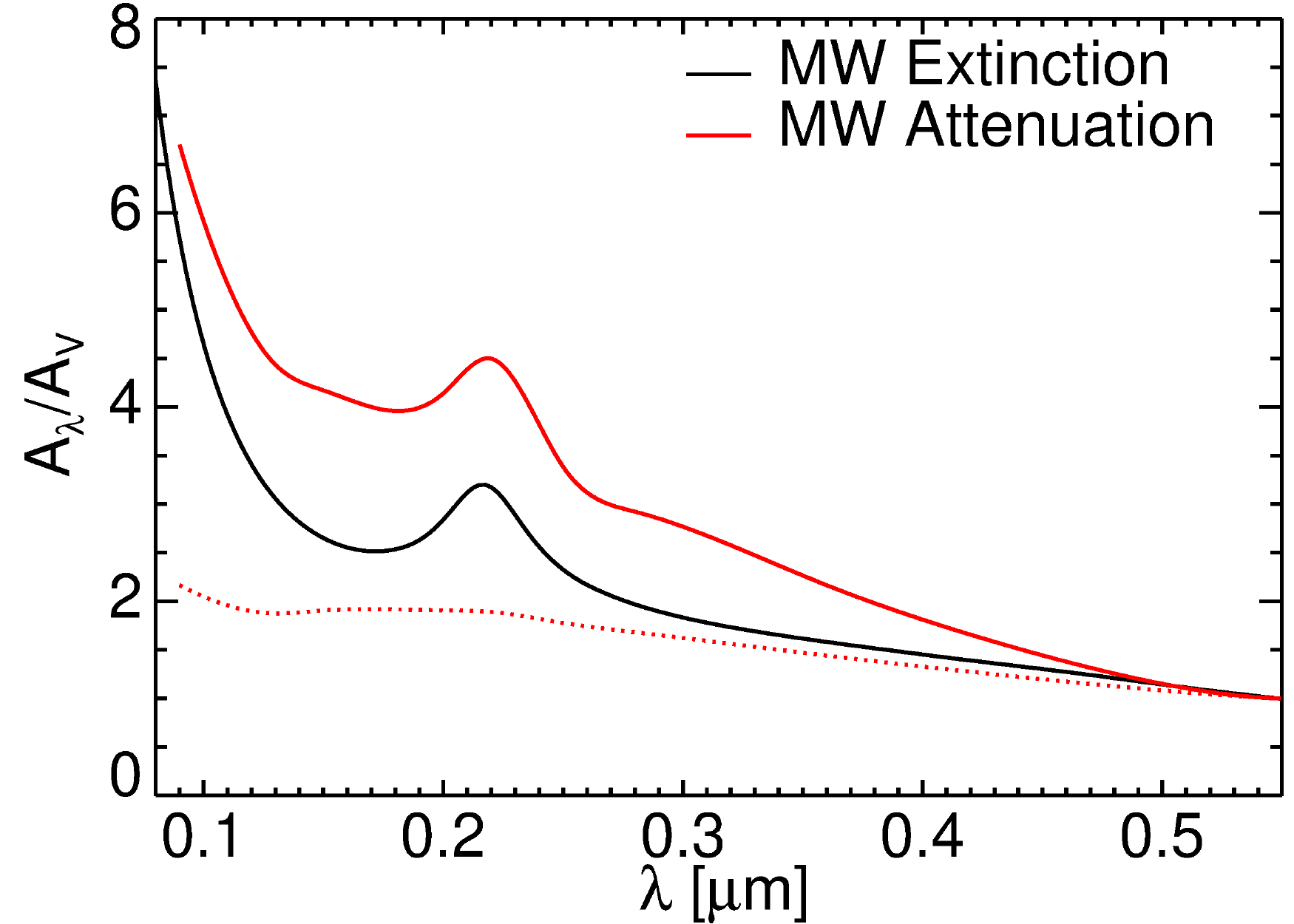}
\caption{The predicted normalised attenuation curve of the Milky Way seen from an outside observer at an intermediate inclination of $56^{\circ}$ (solid red line) and at $90^{\circ}$ (dotted red line). For comparison the average normalised extinction curve of the Milky Way (from Fitzpatrick et al. 1999) is plotted with black line. All curves are normalized to the corresponding values in the V-band.}
\label{fig:atten}
\end{figure}

For this we produced images of the Milky Way as it would be seen by an outside observer  with and without dust, at an inclination of $56^{\circ}$ and $90^{\circ}$. By spatially integrating the apparent and intrinsic emissions  we produced the attenuation curve of the Milky Way as seen at two different orientations. 

\subsubsection{The attenuation curve of the Milky Way at an average inclination}

The attenuation curve of the Milky Way at an average inclination is probably the most interesting to derive, as it can be more readily compared to other curves from the literature, in particular to average attenuation curves derived over populations of galaxies using phenomenological models.

In order to provide an easy access to it we fitted the model of the normalised attenuation curve ($A_{\lambda}/A_{\rm V}$) with the functional form used in Salim et al. (2018), which is a 3rd order polynomial plus a Drude profile (their Eqs. 8 and 9). Using this fit we obtain the  attenuation curve of the Milky Way at an average inclination ($56^{\circ}$):
\begin{eqnarray}
\label{eq:atten_curve}
\kappa_{\lambda}=-5.11+2.10\,{\lambda}^{-1}-0.28\,
{\lambda}^{-2}+0.014\,{\lambda}^{-3}+D_{\lambda}+3.02\\
D_{\lambda}=\frac{1.08\,{\lambda}^2(0.035\,{\mu}m)^2}
{[{\lambda}^2-(0.2175\,{\mu}m)^2]^2{+\lambda}^2(0.035\,{\mu}m)}
\end{eqnarray}
as plotted in Fig.~\ref{fig:atten} with red solid line. The figure also compares this  attenuation curve  with the standard Fitzpatrick extinction curve (whereby both curves have been normalised to their V-band values). One can see immediately that the attenuation curve is steeper than the extinction curve. This is consistent with the predictions from Tuffs et al. (2004) for spiral galaxies, for the inclination and dust opacity range considered in the model curve, and also found by other studies of local Universe star-forming galaxies, including Burgarella et al. (2005), Conroy et al. (2010), Leja et al. (2017),  Salim et al. (2018). Also, the recent studies based on 
radiative 
transfer  calculations found similar trends for M51 (de Looze et al. 2014), M31 (Viaene et al. 2014), and M33 (Williams et al. 2019, Thirlwall et al. 2020). 

The strength of the 2200 \AA\ bump does not seem to vary much between the extinction and the attenuation curve (at $56^{\circ}$). Salim et al. (2018) found that the average attenuation curve of star-forming galaxies exhibit a range of bump strengths, but that they rarely exceed the value of the MW extinction curve. 
 Interestingly, we can now confirm the result for the very Milky Way galaxy. 

\subsubsection{The attenuation curve of the Milky Way as seen edge-on}

The attenuation curve of the Milky Way, as seen by an outside observer at  $90^{\circ}$ inclination, gives insights on the variation of this curve with inclination. One can compare it with the corresponding curve at an average inclination (see solid and dotted red lines in Fig.~\ref{fig:atten}). As expected, the increase in optical depth along the edge-on lines of sight makes the curve flatter, since the Galaxy starts to be more optically thick throughout the whole range of UV wavelengths. The overall effect is that the attenuation curve becomes not only flatter than the attenuation curve at $56^{\circ}$, but even flatter than the extinction curve (see  Fig.~\ref{fig:atten}). The 2200 \AA\ bump completely disappears in this edge-on view, solely as an effect of increased opacity.

\subsection{Comparison of the Milky Way with external galaxies}
\label{sec:comparison}

The Milky Way has been used as our nearest laboratory for studies of galactic
archeology, under the assumption that our own Galaxy is a typical spiral in the
local Universe. But is this true? This question has been raised by different studies, including those trying to understand how representative the MW halo and nearby environment is. Thus, Robotham et al. (2012) analysed the GAMA Galaxy Group Catalogue (G3Cv1) groups (Robotham et al. 2011) drawn from the GAMA survey (Driver et al. 2011) with the aim of understanding how common is to observe a galaxy group with the same characteristics like the MW-LMC-SMC group.

They found that such analogues  are quite rare, occuring with only 0.4 per cent probability. This would indicate that the MW and its close environment is not typical. But, as Bland-Hawthorne \& Gerhard (2016) indicated in their review, the MW is typical in some key respects, but atypical in others. 

\begin{figure}
\centering
\includegraphics[scale=0.15]{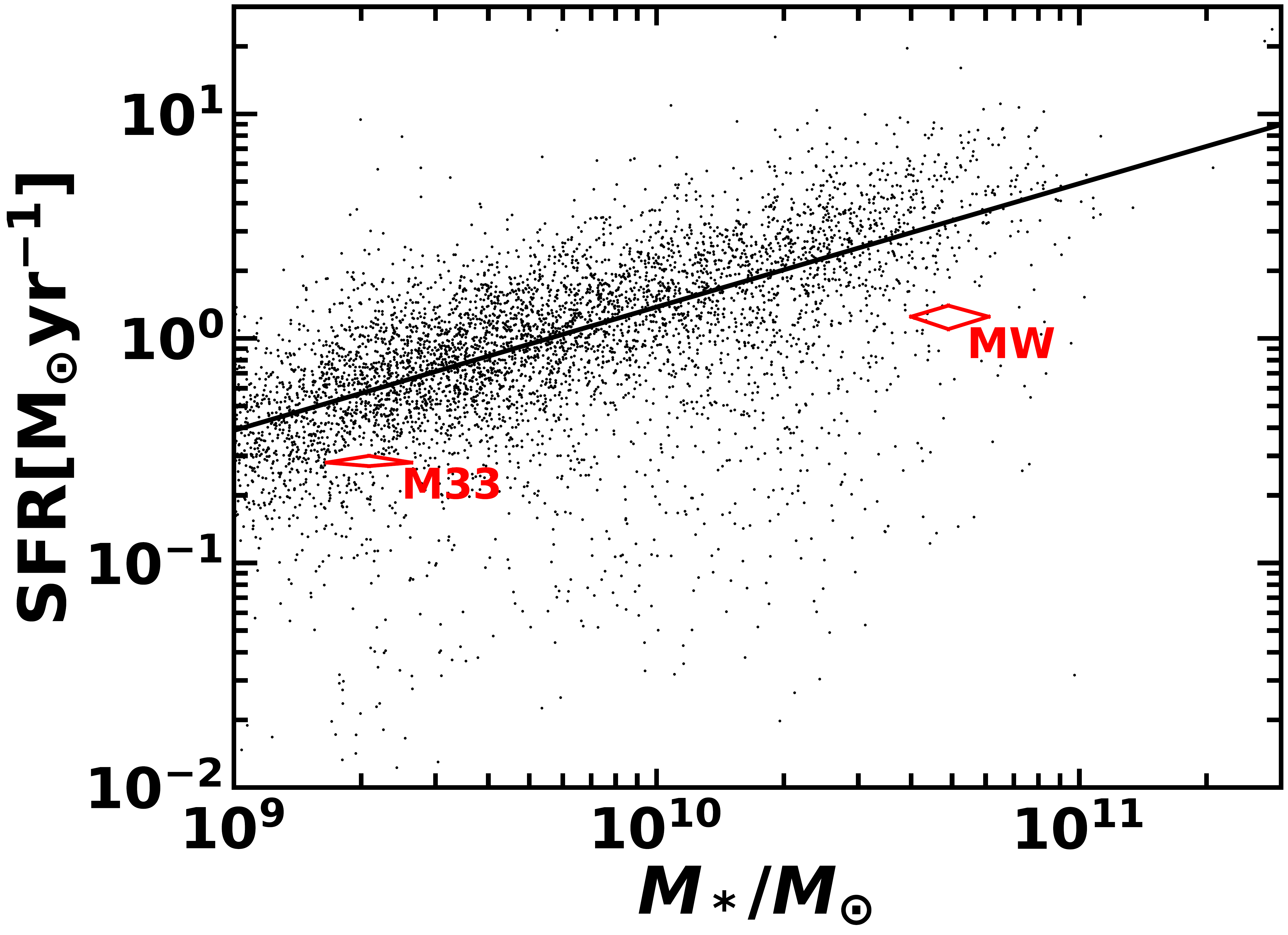}
\caption{The position of the Milky Way in the SFR versus stellar mass relation, as defined by the field reference sample of Grootes et al. (2017). The solid line is the regression fit to a single power-law model given in Table~2 of Grootes et al. (2017).}
\label{fig:gama}
\end{figure}

Having done a multiwavelength SED modelling of the MW it is now
interesting to ask again this question based on the current results. For this we plot the position of the MW in the  star formation
rate vs stellar mass relation (Fig.~\ref{fig:gama}), using as a comparison a volume limited sample of
5202 morphologically selected, disk-dominated galaxies drawn from the GAMA survey (Driver et al. 2011, Hopkins et al. 2013, Liske et al. 2015) by Grootes et al. (2017). The GAMA data were corrected for dust attenuation using the same radiative transfer 
models as used in this paper, under the formalism from Popescu et
al. (2011). SFRs for the GAMA galaxies were derived from the intrinsic i- and g-band photometry following Taylor et al. (2011). The median for each bin in stellar mass is plotted as black solid
symbols, while the main sequence relation is defined by the dashed line. One can
see that the MW lies just below the main sequence of disk galaxies, in the so
called Green Valley. This result shows that the Milky Way is slightly more
quiescent than a typical spiral of the same mass and is consistent
with Licquia et al. (2015), who also finds the Milky Way to lie in the green
valley. Our conclusion is that the MW is not a typical spiral on the \lq\lq blue sequence\rq\rq\,, but a spiral in transition. 

\section{Summary}
\label{sec:summary}

Radiative transfer modelling of the spatial and spectral energy distribution of galaxies is critical in deriving the underlying intrinsic large-scale distributions of stellar emissivity and dust. This type of modelling  has been usually applied to galaxies other than our own, since it is easier to solve the inverse problem for an outside view. In this paper we showed how our models for external galaxies (e.g. Popescu et al. 2011) have been successfully adapted for the inside view of the Milky Way. 

We derived an axi-symmetric model of the Milky Way based on the available all-sky observed maps ranging from the NIR to submm. We used zodiacal-light subtracted maps from COBE, IRAS and PLANCK and limited the comparison between data and models within a strip of fixed size in latitude centred around the Galactic Plane, to avoid contamination from nearby Cirrus clouds.
The main results are as follows:

\begin{itemize}
    \item 
    We derived a total ${\rm SFR}=1.25\pm 0.2 {\rm M}_{\odot}/$yr, of which $1\,{\rm M}_{\odot}/$yr is distributed in a thin stellar disk encompassing the whole extent of the Milky Way, and $0.25\,{\rm M}_{\odot}/$yr represents obscured star formation mostly happening in the inner 4.5\,kpc. The surface density of SFR averaged over the whole Galaxy is  $\Sigma_{\rm sfr}=(2\pm 0.3)\times 10^{-3}{\rm M}_{\odot}\,{\rm yr}^{-1}\,{\rm kpc}^{-2}$. The $\Sigma_{\rm SFR}$ varies across the disk of the Milky Way by 2 orders of magnitude.
    
    \item The stellar mass of the Milky Way is $M_*=4.9\pm0.3 \times 10^{10}\,{\rm M}_{\odot}$.
    \item The specific star-formation rate (averaged over the whole Galaxy) is ${\rm sSFR}=2.6\pm 0.4 \times 10^{-11}$\, yr$^{-1}$. Except for the inner 1\,kpc, the Milky Way has a relatively constant sSFR at around ${\rm sSFR}\simeq 5.\times 10^{-11}{\rm yr}^{-1}$.
    \item The face-on B-band dust opacity distribution has a maximum of $ 1.48\pm 0.1$ at a radial distance of 4.5\,kpc.
    \item The scale-length of the old stellar disk is 
    $h_{\rm s}^{\rm disk}(K)=2.2\pm 0.6$\,kpc and $h_{\rm s}^{\rm disk}(B)=3.2\pm 0.9$\,kpc.
    \item The  scale-height of the old stellar disk is $z_{\rm s}^{\rm disk}(0)=140\pm 20$\,pc, $z_{\rm s}^{\rm disk}(4.5\, {\rm kpc})=170\pm 20$\,pc and $z_{\rm s}^{\rm  disk}(8\, {\rm kpc})=300\pm 20$\,pc.
    \item The  scale-length of the young stellar disk is $h_{\rm s}^{\rm tdisk}=3.2\pm 0.9$\,kpc.
    \item The  scale-height of the young stellar disk is $z_{\rm s}^{\rm tdisk}(0)=50\pm 10$\,pc and $z_{\rm s}^{\rm tdisk}(8\, {\rm kpc})=90\pm 10$\,pc.
    \item We found an inner stellar disk within the inner 4.5 kpc which may be the counterpart of the so-called long-bar of the Milky Way (Hammersley et al. 2000, Cabrera-Lavers et al. 2007, 2008, Benjamin et al. 2005, Wegg et al. 2015).
    \item The   scalelength of the dust disk is $h_{\rm d}=5.2\pm 0.8$\, kpc.
    \item $71\%$ of the dust heating ($F^{\rm dust}$) is powered by the young stellar populations in the thin stellar disk and inner thin stellar disk. Although the young stars dominate the heating mechanisms when integrating over the whole Galaxy, we found that this is not always the case at local scales. The old stellar populations from the bulge dominate the dust heating in the inner 1\,kpc, as well as at some higher vertical distances above the plane. The variation of $F^{\rm dust}$ with vertical position is not monotonic, but has a local minimum/maximum. This is a result of the different vertical distributions of young and old stars and dust opacity effects. 
    \item We predict the  attenuation curve of the Milky Way, as seen by an external (to the Milky Way)  observer (at an average inclination of $i=56^{\circ}$) 
    and present the result in terms of a functional fit (Eq.~\ref{eq:atten_curve}). We find that the slope of the  $i=56^{\circ}$
    MW attenuation curve is steeper than that of the MW extinction curve, with a similar strength of the $2200\,$\AA bump. 
    We also predict the attenuation curve at an edge-on inclination and find  its slope  to be flatter than that of the extinction curve. The bump completely disappears in the $i=90^{ \circ}$ attenuation curve.
    \item The position of the MW in the space defined by the  star-formation rate vs stellar mass relation is slightly below the \lq\lq blue sequence", consistent with the  Milky Way lying in the Green Valley.
\end{itemize}

\section*{Acknowledgements}
We would like to thank the referee, Dr. Emmanuel M. Xilouris for his insightful and helpful comments. Dr. Richard J. Tuffs is acknowledged for critical input to this work.
Cristina C Popescu and Giovanni Natale acknowledge support from a past 
Leverhulme Trust Research Project Grant RPG-2013-418. This research has made use of the NASA/IPAC Infrared Science
Archive, which is operated by the Jet Propulsion Laboratory,
California Institute of Technology, under contract with the National
Aeronautics and Space Administration. 
We acknowledge the use of data provided by the Centre d'Analyse de Données Etendues (CADE), a service of IRAP-UPS/CNRS (http://cade.irap.omp.eu, Paradis et al., 2012).
 Planck data have been used in this paper.
Planck (http://www.esa.int/Planck) is a project of the European
Space Agency (ESA) with instruments provided by two scientific
consortia funded by ESA member states (in particular the lead countries
France and Italy), with contributions from NASA (USA) and
telescope reflectors provided by a collaboration between ESA and
a scientific consortium led and funded by Denmark.

\section{Data Availability}

The data underlying this article are made available at the CDS database
via http://cdsweb.u-strasbg.fr/cgi-bin/qcat?J/MNRAS/.






\appendix
\onecolumn

\section{Error Analysis for the observed surface brightness profiles}
\label{sec:proferrors}

The errors in the surface brightness profiles have three components: calibration errors, background fluctuations and configuration
noise.

The background noise was derived as follows. We first estimated the background in regions of 2 deg in latitude above and below the Galactic Plane Strip and in longitude bins  of 1 deg. For each bin in longitude $s$ we define a background strip $s$ which is further divided in latitude bins $i$. The sky value $\overline{F}_{\rm bg,i}$ for the latitude bin $i$ is then derived as an average of the brightness $F_{\rm n}$ within that bin:

\begin{equation}
\overline{F}_{\rm bg,i} = \frac{1}{N_i} \sum_{n=1}^{N_i}F_{\rm n}
\end{equation}
where N$_i$ is the total number of pixels within the latitude bin $i$. The error in this average is calculated from the pixel-to-pixel variation
within each bin $i$ within the background strip. Thus the pixel-to-pixel variation $\sigma_{\rm bg,i}$ is then:

\begin{equation}
\sigma_{\rm bg,i}=\sqrt{\frac{1}{N_i-1} \sum_{n=1}^{N_i}(F_n-\overline{F}_{\rm bg,i})^2}   
\end{equation}
and the error in the pixel-to-pixel variation for bin $i$ is

\begin{equation}
\epsilon_{\rm bg,i}=\frac{\sigma_{\rm bg,i}}{\sqrt{N_{\rm i}}}
\end{equation}
These uncertainties are then input into the linear function fit which predicts the background within each of the longitude bin $s$ of the Galactic Plane Strip. The end result is a set of background values $F_{\rm bg,fit,i}$ and associated uncertainties $\epsilon_{\rm bg,fit,i}$ for each sampled point $i$ within each background strip $s$. Then, for each strip in latitude for which an averaged $F_{\rm bg,fit, i}$ is calculated, we derive the background errors for that strip by adding in quadrature the uncertainties $\epsilon_{\rm bg,fit,i}$ for each sampled point $i$ within that strip $s$.

\begin{equation}
    \epsilon_{\rm bg,s} =\sqrt{ \sum_{i=1}^{N_s} (\epsilon_{\rm bg,fit,i})^2}
\end{equation}
where $N_s$ is the total number of latitude bins in the strip $s$.
The strips are then mirrored when producing the final averaged profiles, and as such the background error in each averaged strip is found 

by calculating the RMS for each mirrored pair. The background noise in the latitude profiles is calculated following a similar procedure as for the longitude profiles.

Another component of errors in  the average surface brightness profiles is what we call the configuration noise, which arises from deviations of the observed brightness from an assumed axisymmetric distribution. 
The configuration noise was calculated as follows. For the average longitude profiles we consider for each bin in longitude the Q=4 strips in latitude that were used to derive the average: the strip above the plane at the given longitude, the strip above the plane at the corresponding mirrored longitude, the strip below the plane at the given longitude and the strip below the plane at the corresponding mirrored longitude. The average surface brightness $\overline{F}_{\rm gal,q}$ within each strip q = [1,Q] is
given by:

\begin{equation}
    \overline{F}_{\rm gal,q}=\frac{1}{N_q}\sum_{n=1}^{N_q}F_{\rm n}
\end{equation}
where $N_q$ is the total number of pixels within the strip $q$. The average surface brightness over all strips within a bin in
longitude is then:

\begin{equation}
    \overline{F}_{\rm gal}=\sum_{q=1}^{Q} \overline{F}_{\rm gal,q}
\end{equation}
The configuration noise RMS (strip-to-strip variation) $\sigma_{\rm SB,conf}$ is given by:

\begin{equation}
    \sigma_{\rm SB,conf}=\sqrt{\frac{1}{Q-1} \sum_{q=1}^Q (\overline{F}_{\rm gal}-\overline{F}_{\rm gal,q})^2}
\end{equation}
and the configuration error:

\begin{equation}
    \epsilon_{\rm SB,conf}= \frac{\sigma_{\rm SB,conf}}{\sqrt{Q}}
\end{equation}

The configuration noise for the average latitude profiles was calculated in a similar manner. Thus, for each bin in latitude we consider the $Q=4$ strips in longitude that were used to derive the average: the strip with positive longitude\footnote{see our definition of positive and negative longitude from Sect.~\ref{sec:data}} at the given latitude, the strip with positive longitude at the corresponding mirrored latitude, the strip with negative longitude at the given latitude and the strip at negative longitude and the corresponding mirrored latitude. The errors for the average latitude profiles are then given by the same formulas from Eqns 1-4.

The total errors in the averaged surface brightness profiles have been derived using:

\begin{equation}
    \epsilon_{SB_\nu}=\sqrt{\epsilon_{\rm cal}^2+\epsilon_{\rm SB,bg}^2+\epsilon_{\rm SB,conf}^2}
    \label{eq:total_error}
\end{equation}
whereby the first term is independent of longitude/latitude, while the second and third terms are longitude/latitude dependent.

\section{The stellar luminosity
and the dust mass}
\label{sec:formulae}

The spatial integration of the disk emissivity (Eq.~\ref{eq:model}) up
to the truncation radius $R_{\rm t}$ and the truncation height $z_{\rm
  t}$, where $z_{\rm t} >>z_{\rm i}$,
is given by:

\begin{equation}
\label{eq:integral}
I  =  4\pi A_0\left[\left(1+\frac{\chi}{2}\right) \frac{R_{\rm in}}{3} \exp\left(-\frac{R_{\rm
        in}}{h_{\rm i}}\right) +  h_{\rm i}^2 T_{\rm R}\right]z_{\rm
  i}T_{\rm z}
\end{equation}
where
\begin{equation}
\label{eq:TR}
T_{\rm R}  =  \exp\left(-\frac{R_{\rm in}}{h_{\rm i}}\right)-\exp\left(-\frac{R_{\rm
      t}}{h_{\rm i}}\right)+\frac{R_{\rm in}}{h_{\rm
      i}} \exp\left(-\frac{R_{\rm in}}{h_{\rm
          i}}\right)-\frac{R_{\rm t}}{h_{\rm
          i}} \exp\left(-\frac{R_{\rm t}}{h_{\rm
          i}}\right)
\end{equation}
and
\begin{equation}
\label{eq:Tz}
T_{\rm z}  =  1-\exp\left(-\frac{z_{\rm
    t}}{z_{\rm i}}\right) 
\end{equation}

\noindent
In the case of a  stellar disk component, 
Eqs.~\ref{eq:integral},\ref{eq:TR},\ref{eq:Tz} provide its spatially
integrated stellar luminosity, by taking $A_0$ to be the central
volume luminosity density $L_0$, and i=\lq s\rq. Thus the stellar
disk luminosity is
\begin{equation}
\label{eq:lum}
L=I(L_0,h_{\rm s},z_{\rm s}),
\end{equation} where $h_{\rm s}$,$z_{\rm s}$ are the
scalelength and height of that disk.

In the case of a dust component
Eqs.~\ref{eq:integral},\ref{eq:TR},\ref{eq:Tz}  provide its dust mass,
by taking $A_0$ to be the central volume density of dust
\begin{equation}
\label{eq:density}
\rho_{\rm c}=\tau_{\rm c}/(2\kappa z_{\rm d})
\end{equation}
 and i=\lq d\rq, where $\tau_c$ is the
central face-on dust opacity and $\kappa$ is the mass extinction
coefficient. Thus the dust mass is 
\begin{equation}
\label{eq:mass}
M_{\rm d}=I(\rho_{\rm c},h_{\rm d},z_{\rm d}),
\end{equation} where $h_{\rm d}$,$z_{\rm d}$
are the
scalelength and height of that dust disk.

\section{The error analysis for the model}
\label{sec:errors}

\begin{figure*}
\centering
\includegraphics[scale=0.49]{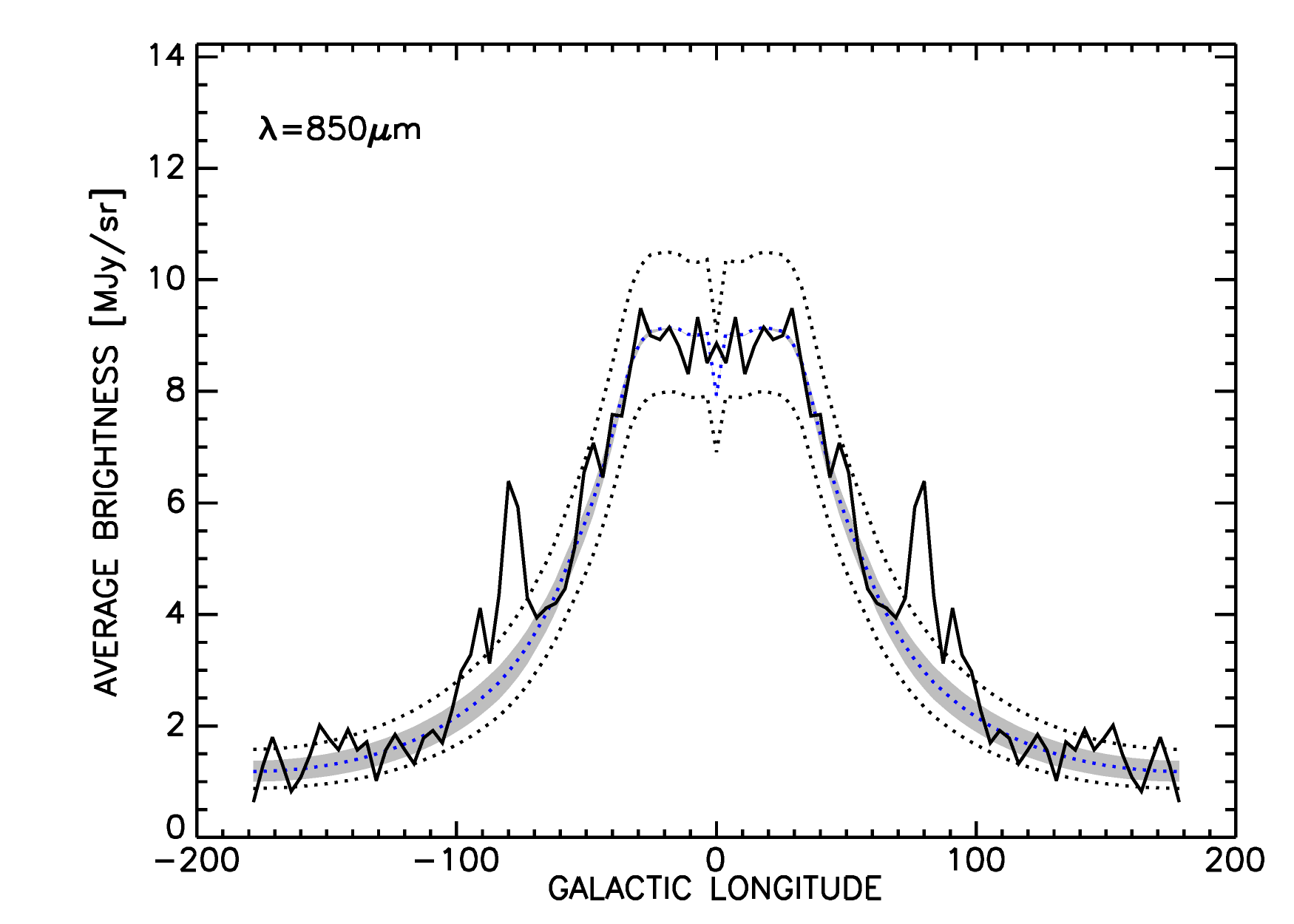}
\includegraphics[scale=0.49]{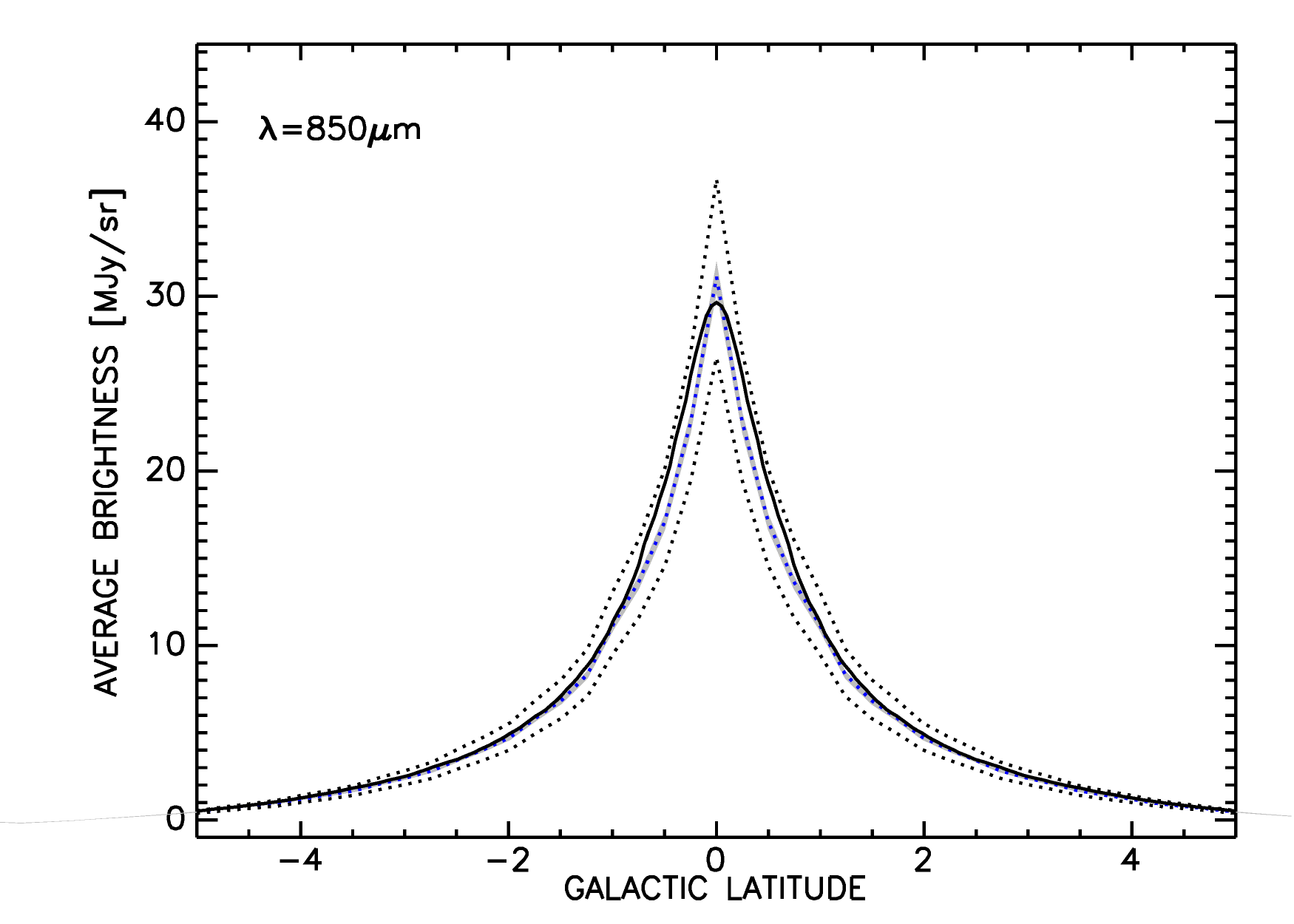}
\caption{Variation in the average longitude and latitude model
  profiles (dotted lines) of surface brightness at $850\,{\mu}$m 
due to $15\%$ variation in the $h_{\rm d}$. The corresponding observed profiles are plotted with a solid line. The shaded area
represents the variation in the models after the same
change in $h_{\rm d}$ but this time accompanied by a change in the
$\tau^{\rm f}(B)$ parameter, such that the centre region of the 
$850\,{\mu}$m longitude profile is fitted. This is equivalent to the
conditional probability analysis conducted to find errors in 
$h_{\rm d}$ and $\tau^{\rm f}(B)$.} 
\label{fig:error_850}
\end{figure*}

The uncertainties in the main geometrical parameters of the model
(those that are constrained from data) were derived by looking at the
departure from the best-fitting model of only one parameter at a
time, at the wavelength at which the parameter 
was optimized. For example, for the scalelength of the thick dust disc, $h_{\rm d}$, we show in Fig.~\ref{fig:error_850} how the fit to the averaged 
longitude and latitude profiles of surface brightness changes 
for a change in $h_{\rm d}$ (dotted lines) that corresponds to the adopted error in $h_{\rm d}$. Because the large
variation in amplitude was compensated for in the optimization by a
subsequent variation in the amplitude parameter,  $\tau_{\rm B}^{\rm f}$, we also show the variation after the profiles
were rescaled to fit the central flat part of the longitude profiles
(shaded area). The shaded area is then taken to represent the
uncertainty in the model fit.

\begin{figure*}
\centering
\includegraphics[scale=0.49]{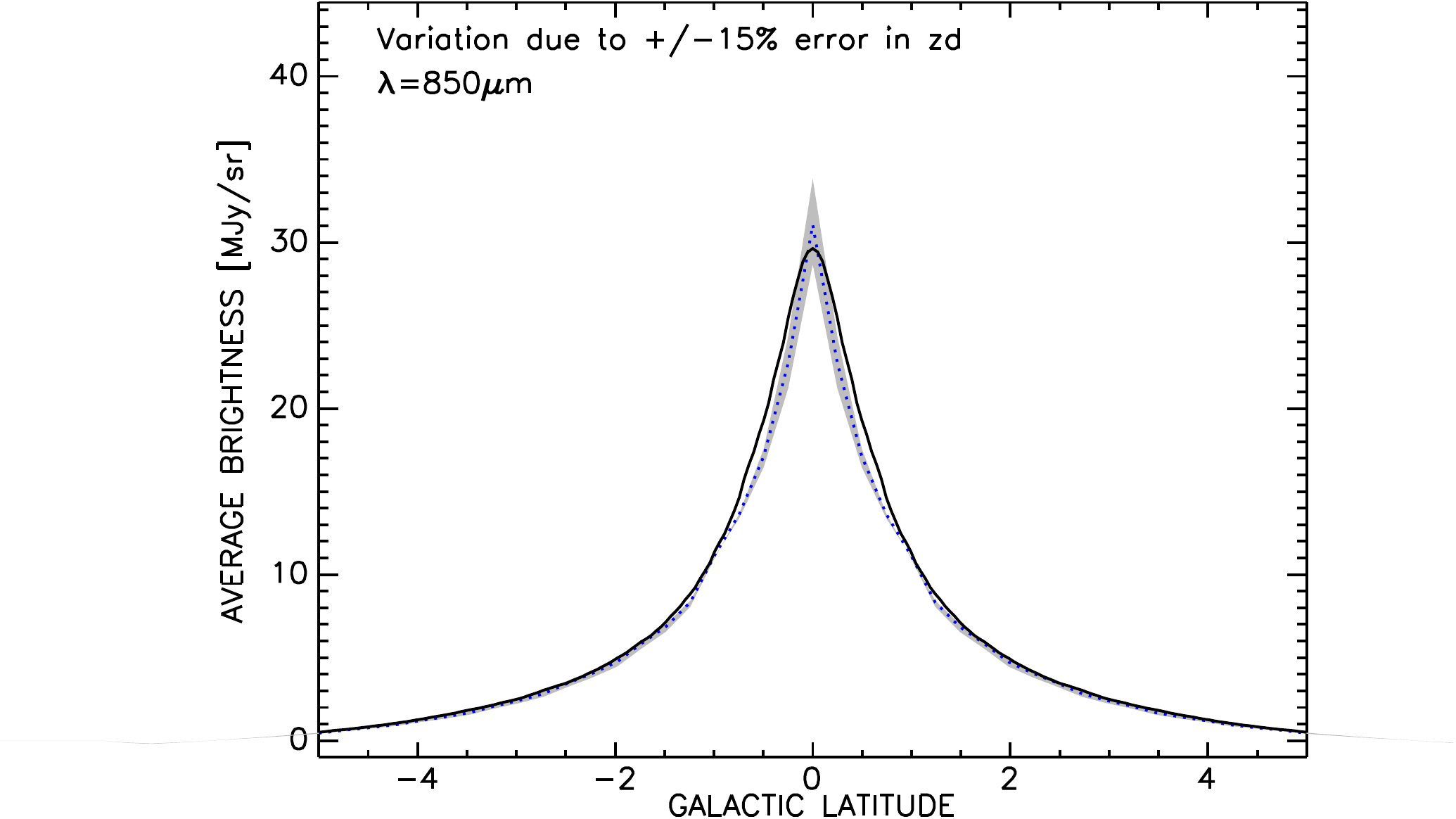}
\caption{Variation in the average latitude model
  profiles (shaded area) of surface brightness at $850\,{\mu}$m 
due to $15\%$ variation in the $z_{\rm d}$ and a change in the
$\tau^{\rm f}(B)$ parameter, such that the centre region of the 
$850\,{\mu}$m longitude profile is fitted. The corresponding observed profiles are plotted with a solid line.} 
\label{fig:error_lat_850}
\end{figure*}

\begin{figure*}
\centering
\includegraphics[scale=0.49]{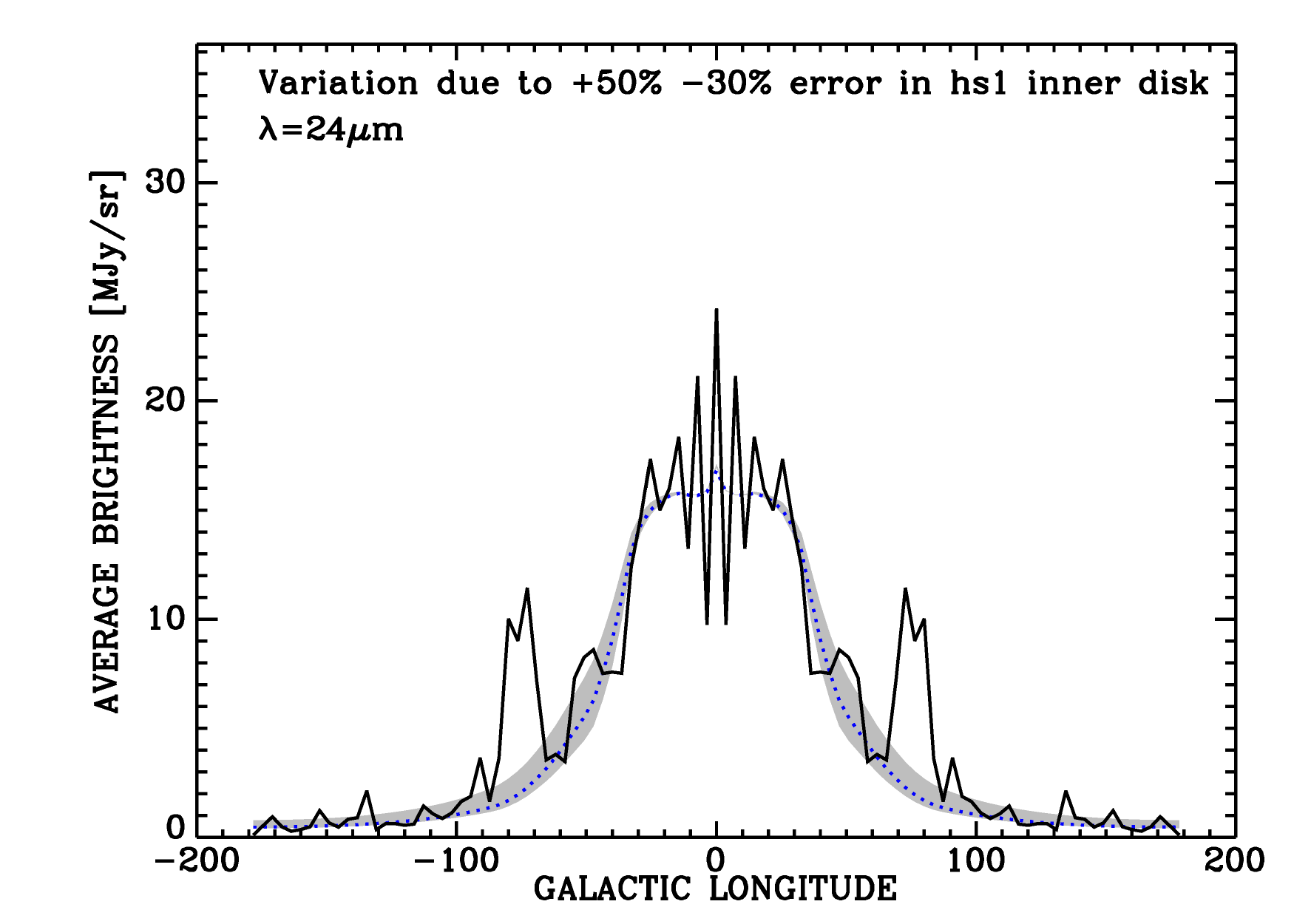}
\includegraphics[scale=0.49]{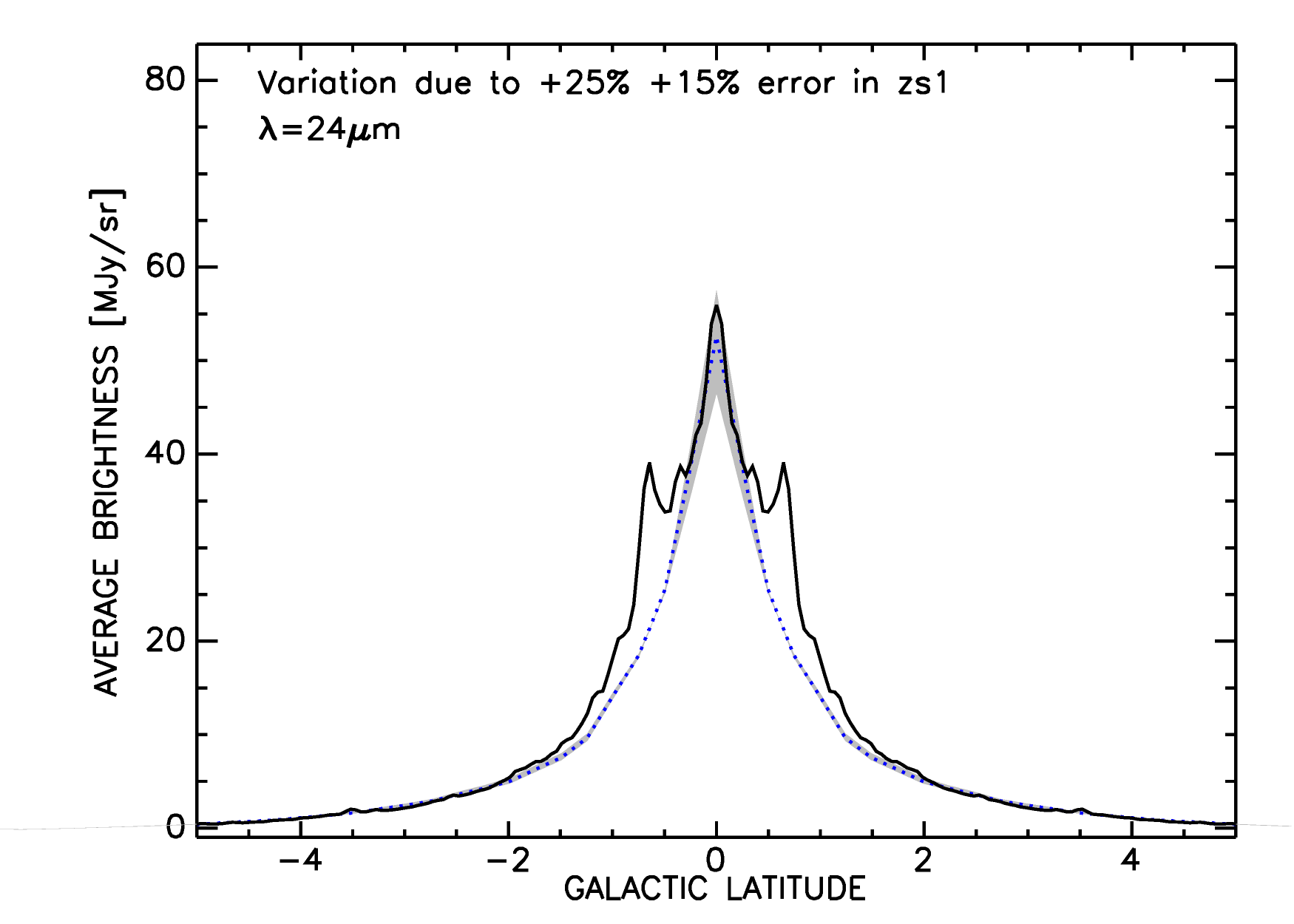}
\caption{Left: Variation in the average longitude model
  profiles (shaded area) of surface brightness at $24\,{\mu}$m 
due to $+50\% -30\%$ variation in the $h_{\rm s}^{\rm in-tdisk}$
and a change in the
$\rm SFR^{\rm in-tdisk}\times F^{\rm \, in-tdisk}$ parameter, such that the centre region of the 
$24\,{\mu}$m longitude profile is fitted. Right: Variation in the average latitude model
  profiles (shaded area) of surface brightness at $24\,{\mu}$m 
due to $+25\% -15\%$ variation in the $z_{\rm s}^{\rm in-tdisk}$
and a change in the
$\rm SFR^{\rm in-tdisk}\times F^{\rm \, in-tdisk}$ parameter, such that the centre region of the 
$24\,{\mu}$m longitude profile is fitted. The corresponding observed profiles are plotted with a solid line.} 
\label{fig:error_24}
\end{figure*}

\begin{figure*}
\centering
\includegraphics[scale=0.49]{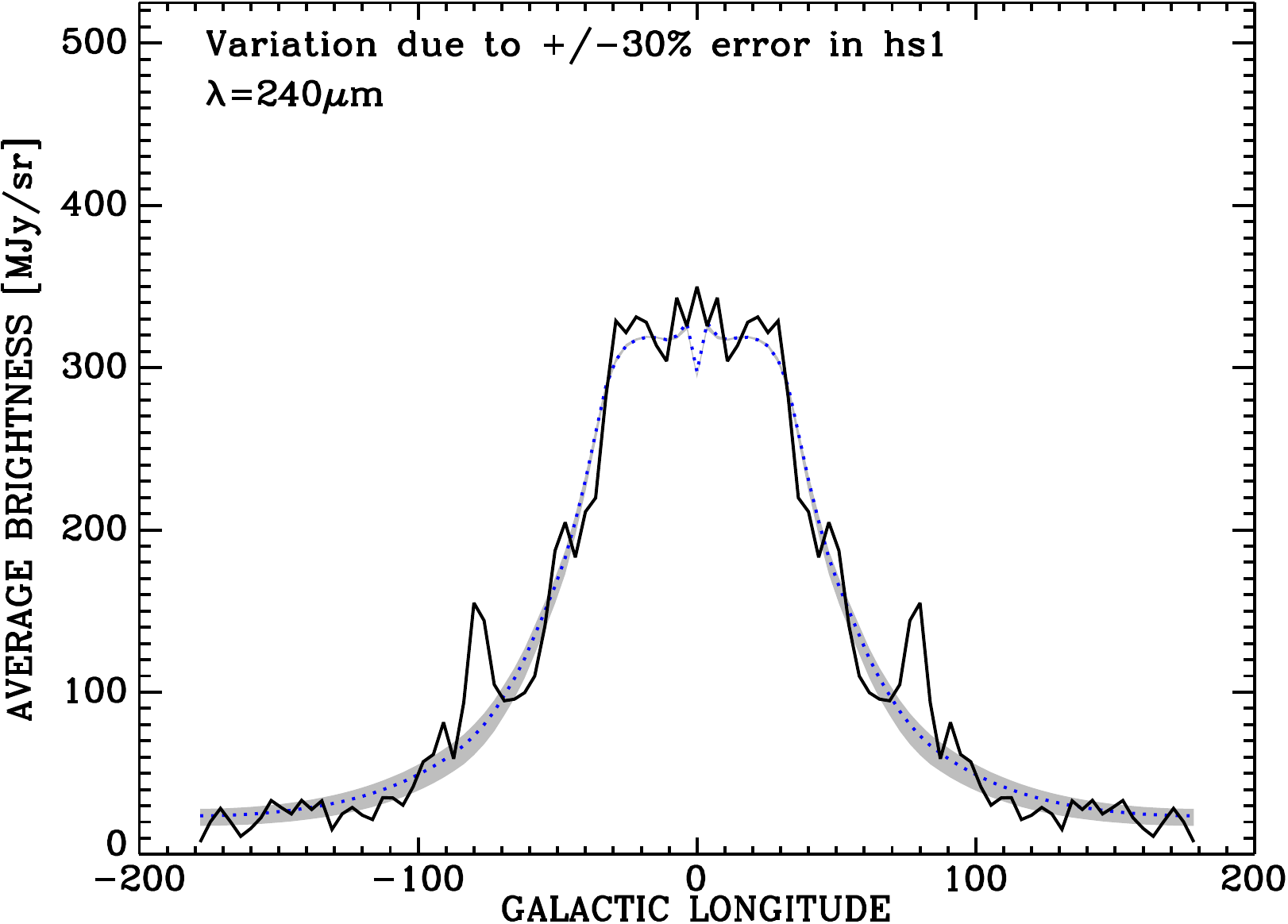}
\caption{Variation in the average longitude model
  profiles (shaded area) of surface brightness at $240\,{\mu}$m 
due to $30\%$ variation in the $h_{\rm s}^{\rm tdisk}$
and a change in the
$\rm SFR^{\rm tdisk}\times(1-F)^{\rm \, tdisk}$ parameter, such that the centre region of the 
$240\,{\mu}$m longitude profile is fitted.} 
\label{fig:error_240}
\end{figure*}

\begin{figure*}
\centering
\includegraphics[scale=0.49]{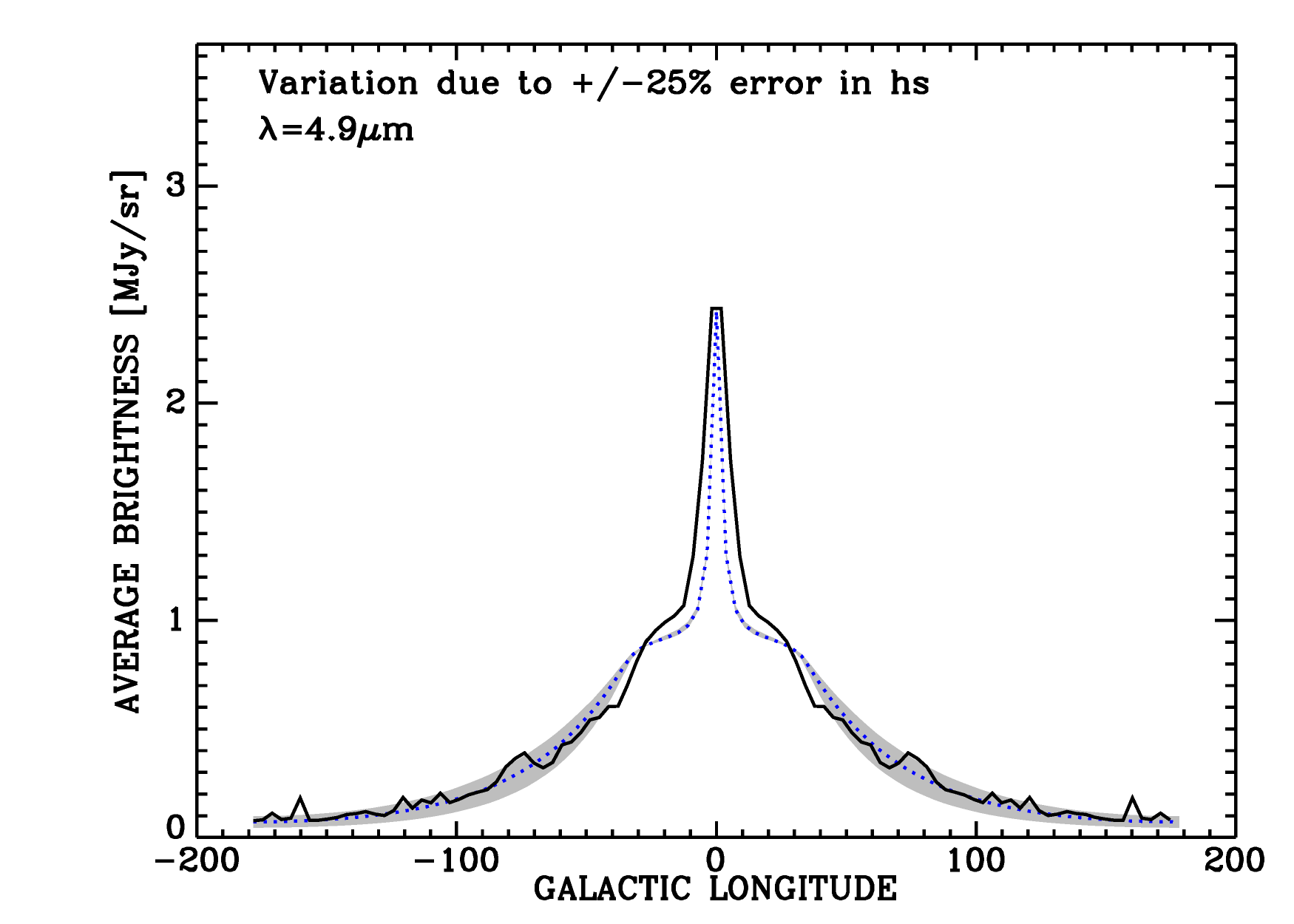}
\includegraphics[scale=0.49]{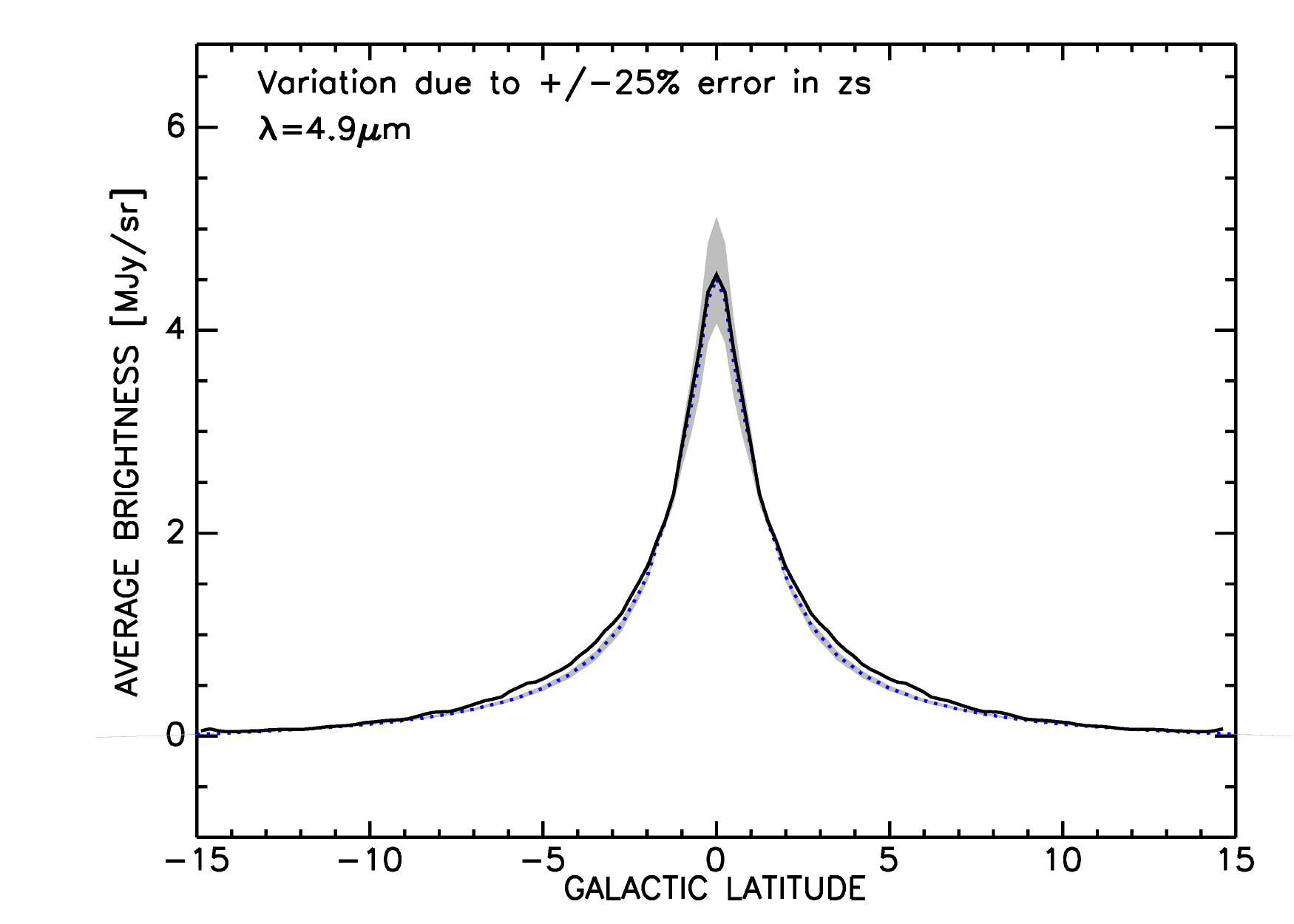}
\caption{Left: Variation in the average longitude model
  profiles (shaded area) of surface brightness at $4.9\,{\mu}$m 
due to $25\%$ variation in the $h_{\rm s}^{\rm disk}({\rm M})$
and a change in the
$L^{\rm disk}({\rm M})$ parameter, such that the centre region of the 
$4.9\,{\mu}$m longitude profile is fitted. Right: Variation in the average latitude model
  profiles (shaded area) of surface brightness at $4.9\,{\mu}$m 
due to $25\%$ variation in the $z_{\rm s}^{\rm disk}({\rm M})$
and a change in the
$L^{\rm disk}({\rm M})$ parameter, such that the centre region of the 
$4.9\,{\mu}$m longitude profile is fitted. The corresponding observed profiles are plotted with a solid line.} 
\label{fig:error_4.9}
\end{figure*}

In a similar way we show in Figs.~\ref{fig:error_lat_850},
\ref{fig:error_24}, \ref{fig:error_240}, \ref{fig:error_4.9} the
variation in the average longitude or latitude profiles due to the
variation of the following pairs of parameters:
$z_{\rm d}$ and $\tau^{\rm f}(B)$, $h_{\rm s}^{\rm in-tdisk}$ and
$\rm SFR^{\rm in-tdisk}\times F^{\rm \, in-tdisk}$, $z_{\rm s}^{\rm
  in-tdisk}$ and $\rm SFR^{\rm in-tdisk}\times F^{\rm \, in-tdisk}$, $h_{\rm
  s}^{\rm tdisk}$ and $\rm SFR^{\rm tdisk}\times(1-F^{\rm \, tdisk})$,
$h_{\rm s}^{\rm disk}({\rm M})$ and $L^{\rm disk}({\rm M})$, $z_{\rm s}^{\rm
  disk}({\rm M})$ and $L^{\rm disk}({\rm M})$, at the corresponding wavelength where the pair of parameters each parameter was optimised.
  
The goodness of the fit to the observed average surface brightness profiles was  quantified through a chi-squared calculation 
at the key wavelengths where the model is optimized (850, 24, 240, 4.9\,$\mu$m): 

\begin{align}
    {\rm chi}_{\lambda}^2&=\sum\limits_{i=1}^{N}\frac{({O}_{\rm i} - {M}_{\rm i})^2}{\varepsilon_{{\rm SB,i}}^{2}}\label{eqn:chisq}\\ 
    {\rm chi}^2_{r,\lambda}&=\frac{{\rm chi}_{\lambda}^2}{N}\label{eqn:chisqr}\\
\end{align}
where $N$ is the number of bins in the latitude or longitude profile, $O_{\rm i}$ and $M_{\rm i}$ are the averaged surface brightnesses for the bin $\rm i$ of  the observed and modelled longitude or latitude profiles, respectively, and $\varepsilon_{\rm SB, i}$ is the error for the bin $\rm i$ of the profile, as derived using Eqn.~\ref{eq:total_error}.    The corresponding reduced chi-squared ${\rm chi^2_r}$ are listed in Table~\ref{tab:chi_r}.
The reduced chi-squared value for the model across all wavelengths 
 is ${\rm chi}_{\rm r}^2=2.29$.

\begin{table}
\caption{The ${\rm chi}^2_{\rm r}$ values for the best-fitting model and the upper and lower error models at the wavelengths where the model was optimized. The table is organised as follows: Column 1 gives the pair of parameters that were constrained from a specific wavelength; Column 2 gives the wavelength where the pair of parameters from Column 1 was optimized; Column 3 gives the type of profile that constrains the pair of parameters, being either average longitude profile (long) or average latitude profile (lat); Column 4 gives the $\rm chi^2_r$ for the best fit model; Column 5 gives the $\rm chi^2_r$ for the upper error model; Column 6 gives the $\rm chi^2_r$ for the lower error model.}
\label{tab:chi_r}
\begin{tabular}{llllll}
\hline
parameter & $\lambda$ [$\mu$m] & profile & Best & e+ & e- \\
\hline
$h_{\rm d}$, $\tau^{\rm f}(B)$ & 850 & long & 1.28 & 3.15 & 4.79\\
$z_{\rm d}$, $\tau^{\rm f}(B)$ & 850 & lat & 0.53 & 2.25 & 3.50\\
$h_{\rm s}^{\rm in-tdisk}$, ${\rm SFR}^{\rm in-tdisk}\times F^{\rm \,in-tdisk}$ & 24 & long & 3.5 & 7.78 & 12.38\\
$z_{\rm s}^{\rm in-tdisk}$, ${\rm SFR}^{\rm in-tdisk}\times F^{\rm \,in-tdisk}$ & 24 & lat & 0.68 & 3.2 & 1.99\\
$h_{\rm s}^{\rm tdisk}$, ${\rm SFR}^{\rm tdisk}\times (1-F^{\rm \,tdisk})$ & 240 & long & 2.18 & 4.72 & 6.17\\
$h_{\rm s}^{\rm disk}({\rm M})$, $L^{\rm disk}$ & 4.9 & long & 2.74 & 5.14 & 8.6\\
$z_{\rm s}^{\rm disk}({\rm M})$, $L^{\rm disk}$ & 4.9 & lat & 0.28 & 0.83 & 2.35\\
\hline
\end{tabular}
\end{table}


\bsp	
\label{lastpage}
\end{document}